\documentclass[12pt,a4paper]{article}
\usepackage[numbers,sort&compress]{natbib}
\usepackage{amsmath,amssymb,tabularx}
\usepackage{graphicx}
\usepackage[title]{appendix}
\usepackage{verbatim}
\usepackage{titlesec}
\titlelabel{\thetitle.\quad}
\usepackage{bm}
\usepackage{subcaption}
\usepackage{dsfont}
\usepackage{color}
\usepackage{physics}
\usepackage{slashed}
\usepackage[svgnames]{xcolor}
\usepackage{hyperref}
\usepackage{hyperref}
\hypersetup{
	colorlinks,
	citecolor=blue,
	filecolor=blue,
	linkcolor=blue,
	urlcolor=blue
}
\usepackage{caption}	
\usepackage{microtype}
\usepackage{ctable}
\newcommand{\bx}{\textbf{x}}
\newcommand{\by}{\textbf{y}}
\newcommand\varpm{\mathbin{\vcenter{\hbox{%
				\oalign{\hfil$\scriptstyle+$\hfil\cr
					\noalign{\kern-.3ex}
					$\scriptscriptstyle({-})$\cr}%
}}}}
\newcommand\varmp{\mathbin{\vcenter{\hbox{%
				\oalign{$\scriptstyle({+})$\cr
					\noalign{\kern-.3ex}
					\hfil$\scriptscriptstyle-$\hfil\cr}%
}}}}
\newcommand{\nn}{\nonumber}
\allowdisplaybreaks
\newcommand{\bea}{\begin{eqnarray}}
\newcommand{\eea}{\end{eqnarray}}
\newcommand{\ba}{\begin{align}}
\newcommand{\ea}{\end{align}}
\newcommand{\bp}{\textbf{p}}
\newcommand{\bq}{\textbf{q}}
\newcommand{\ap}{|\bp|}

\newcommand{\bn}{\textbf{n}}
\newcommand{\A}{\tilde{A}}

\begin{document}
\title{
\begin{flushright}
\ \\*[-80pt] 
\begin{minipage}{0.2\linewidth}
\normalsize

HUPD-2411 \\*[5pt]
\end{minipage}
\end{flushright}
{\Large 
The probability for chiral oscillation of Majorana neutrino in Quantum Field Theory
\\*[5pt]} }
\author{\normalsize
\centerline{
Takuya Morozumi$^{1,2 }$ 
\footnote{morozumi@hiroshima-u.ac.jp},
\ Tomoharu Tahara $^{1}$
\footnote{t-tomoharu@hiroshima-u.ac.jp}
} \\ \normalsize
\centerline{
}
\\*[10pt]
\centerline{
\begin{minipage}{\linewidth}
\begin{center}
$^1${\it \small
Physics Program, Graduate School of Advanced Science and Engineering, Hiroshima University,\\ Higashi-Hiroshima 739-8526, Hiroshima, Japan} \\*[5pt]
$^2${\it \small
Core of Research for the Energetic Universe, Hiroshima University, \\ Higashi-Hiroshima 739-8526, Hiroshima, Japan} \\*[5pt]
\end{center}
\end{minipage}}
\\*[50pt]}
\date{
\centerline{\small \bf Abstract}
\begin{minipage}{0.9\linewidth}
\medskip
\medskip
\small
We  derive the probability for chiral oscillation of Majorana neutrinos based on quantum field theory. 
   Since the Hamiltonian under the Majorana mass term does not conserve lepton number, the eigenstates of lepton number change continuously over time. Therefore, the transition amplitude is described by the inner product of the eigenstates of lepton number at the time of the neutrino production and the detection. With the Bogoliubov transformation,  we successfully relate the lepton number eigenstates at different times.  This method enables us to understand the time variation of lepton number induced by chiral  oscillations in terms of transition probabilities.  We also present the physical picture that emerges through the Bogoliubov transformation.
\end{minipage}
}
\begin{titlepage}
\maketitle
\thispagestyle{empty}
\end{titlepage}
\counterwithin*{equation}{section}
\renewcommand\theequation{\thesection.\arabic{equation}}
\section{Introduction}\label{sec:introduction}
There are two types of neutrino oscillations: flavor oscillation \cite{Maki:1962mu} 
and neutrino anti-neutrino oscillation \cite{Pontecorvo:1957qd, Schechter:1980gk}.
Flavor oscillation is a  phenomenon in which a certain flavor neutrino (either $\nu_e$, $\nu_{\mu}$, or $\nu_\tau$) is produced and observed as a different flavor neutrino through time. On the other hand, neutrino anti-neutrino oscillation
 is a phenomenon characteristic of Majorana neutrinos, in which neutrinos transform into  anti-neutrinos due to the time evolution \cite{Giunti:2007ry, Beuthe:2001rc, Fukugita:1993fr, Fukugita:2003en}.
 Neutrino and anti-neutrino oscillation  has been predicted by \cite{Pontecorvo:1957qd} where the oscillation occurs in the framework of the two chiral neutrinos $\nu_L$ and $\nu_R$ 
with Majorana mass terms as well as Dirac mass term. The quantum mechanical oscillation formula for $\nu_L$ and $(\nu_R)^c$ is given for this model in \cite{Fukugita:1993fr, Fukugita:2003en}. In Pontecorvo's
neutrino and anti-neutrino oscillation is not suppressed by the mass of the neutrino, since it is the transition among the fields with the same chirality. In the context of the Majorana neutrino of $\nu_L$ without $\nu_R$, the oscillation from the anti-electron neutrino to muon neutrino has been discussed with the propagation of massive Majorana neutrinos \cite{Schechter:1980gk} for the relativistic case. 
In this work, we focus on the Majorana neutrino and a phenomenon caused by the Majorana mass term.  
 It leads to the transition among the states with different lepton numbers of neutrinos . 
In contrast to the
framework of  Pontecorvo's neutrino and anti-neutrino oscillation,
The effect becomes significant  when neutrinos carry small momentum compared with 
their rest mass.  The standard oscillation formula for the relativistic neutrino cannot be applied to this case. Since the chiral oscillation\footnote{We use the term "chiral oscillation" as the oscillation phenomena observed in the probability of the event caused by Majorana mass term. In Ref.\cite{Kimura:2021qlh,Bittencourt:2022hwn,Bittencourt:2024yxi}, chiral oscillations in the case of  Dirac masses are discussed. }\cite{Kimura:2021qlh,Bittencourt:2022hwn,Bittencourt:2024yxi} 
caused by Majorana mass term is always accompanied
by the change of the lepton number, we define the time-dependent
transition probabilities among the states with
different lepton numbers.
The Heisenberg operator for the lepton number is introduced in \cite{Adam:2021qiq, SalimAdam:2021suq}.  The lepton number operator is time-dependent and its eigenstates also depend on time.
Then one can define the transition amplitude as an inner product between the state at the time of production and that at the time of detection.
 Each state is chosen as the eigenstate of the lepton number operator at the corresponding  time. 
We introduce the Bogoliubov transformation \cite{Bogolyubov:1958km, Morozumi:2022mqh}
to relate the creation and annihilation operators defined at different times and then one can easily compute the inner product among the states created.

 As the first application of our framework,
we study the simplest system of a one-flavor Majorana neutrino. The time-dependent chiral transition probability is derived.
By using the quantum field theory, we have developed a theoretical framework  that can be applied to both relativistic and 
non-relativistic neutrinos.    The effect of the Majorana mass term is important for the latter.

The outline of this paper is as follows. In section \ref{sec:Hamiltonian}, we introduce the Hamiltonian for the single  Majorana neutrino. We carefully exclude the momentum zero mode
for the Majorana neutrino and quantize the system.  The field is expanded  in terms of  the creation and annihilation operators with the definite lepton number.  Section \ref{sec: Bogoliubov transformation} focuses on deriving  the time evolution of the operators. Using the Bogoliubov transformation, we also show  relations among the eigenstates of the lepton number operator at both times of production and detection.  In section \ref{sec:Probability}, we derive oscillation probabilities based on the time evolution of the operators and eigenstates. 
In section \ref{sec:Conclusion},  we discuss the physical implication of our result which leads to the new interpretation for the lepton number changing chiral oscillation.
  We also show how the expectation value of the lepton number evaluated in 
 \cite{Adam:2021qiq, SalimAdam:2021suq, Morozumi:2022mqh} is related to the probabilities in the present work. 
In Appendix \ref{sec:appendix zeromode},  the derivation of the Hamiltonian and anti-commutation relations of the field operators is given. 
In Appendix \ref{sec:Blasone}, we explain the difference of probabilities between the present 
 approach and the perturbative approach in Ref.\cite{Blasone:2024zsn}.  In Appendix \ref{sec:appendix proof}, we show that the neutrino and anti-neutrino oscillation does not occur for one-flavor case.
\section{Hamiltonian }\label{sec:Hamiltonian}
To quantize the Majorana field \cite{Fukugita:1993fr,Fukugita:2003en,Case:1957zza}, the standard approach is to introduce the creation and annihilation operators for massive Majorana field. However this approach is not suitable for the purpose
to compute the transition amplitude among the states with definite lepton numbers. This is because the one-particle mass eigenstate obtained by applying the creation operator on the time-invariant vacuum does not carry the definite lepton number.   In our approach,
the creation and annihilation
operators are chosen in such a way that the one-particle state has the definite lepton number.  This is achieved by expanding the field operator with massless plane wave spinors and creation and annihilation
operators associated with them. The corollary of
introducing massless spinors is that the time evolution of the operators becomes complex and
the vacuum is time-dependent. On the other hand, the lepton number operator is simply written as the difference of the number operators for neutrino and anti-neutrino as in Eq.(\ref{LeptonNumber}).
As one can not express the zero momentum mode of massive field with the massless spinors,  we need to exclude the zero mode.   
 If we keep the zero mode, one must attribute the mass parameter to operators for the zero mode and the lepton number operator cannot be simply expressed by the difference 
of the number operators for neutrino and anti-neutrino \cite{SalimAdam:2021suq}.
 Below, we show how to exclude the zero mode consistently with the time evolution of operators.
The results imply that one can construct the Hilbert space without the Fock space and the vacuum for the zero mode.

We begin with the path-integral expression of the action for a Majorana neutrino in the single-flavor case,
\begin{align}
 \int d\eta d\eta^{\dagger}\int d\xi_0 d\xi_0^\dagger e^{iS'[\eta,\xi_0]}
= \int\int d\eta d\eta^{\dagger}\delta(\eta_0)\delta(\eta_0^\dagger)e^{iS[\eta]},
\label{pathintegral}
\end{align} 	
where in the right-hand side of Eq.(\ref{pathintegral}), $S[\eta]$ is an action for a single flavor Majorana neutrino with mass $m$ in terms of two-component chiral field $\eta$,
\begin{align}
S[\eta] = \int d^4x \mathcal{L},\quad
\mathcal{L} = \eta^{\dagger}(i\bar{\sigma}^{\mu}\partial_{\mu})\eta -\frac{m}{2}(- \eta^{\dagger}i\sigma_2\eta^{\dagger} + \eta i\sigma_2\eta).
\end{align}
$\eta_0$ denotes the zero mode of $\eta$ defined as,
\begin{align}
\eta_0(t) = \frac{1}{V}\int d^3\bx \eta(t,\bx),
\end{align}
where $V$ denotes the space volume and is defined by $V=(2\pi)^3\delta^{(3)}(\bp=0)$. 
 In the path-integral expression of the right-hand side,
the delta functions $\delta(\eta_0)$ and  $\delta(\eta^\dagger_0)$ remove  zero modes from the path-integral and the action.
In the left-hand side of Eq.(\ref{pathintegral}), 
we express the delta function using the following formula,
\begin{align}
\delta(\eta_0) \delta(\eta_0^\dagger)  &= \int d\xi_{0}^\dagger  d\xi_{0}  e^{(\xi_0^\dagger\eta_0
-\eta_0^\dagger \xi_0)}.
\end{align}
Then the action $S'[\eta,\xi_0]$ is given as,
\begin{align}
S'[\eta,\xi_0] = S[\eta] -i\int dt(\xi_0^\dagger\eta_0-\eta_0^\dagger \xi_0)
=\int d^4 x \mathcal{L}',
\end{align}
where the Lagrangian density $\mathcal{L}'$ is given by,
\begin{align}
\mathcal{L}' 
&= \eta^{\dagger}(i\bar{\sigma}^{\mu}\partial_{\mu})\eta -\frac{m}{2}(- \eta^{\dagger}i\sigma_2\eta^{\dagger} + \eta i\sigma_2\eta)-\frac{i}{V}(\xi_0^\dagger\eta_0-\eta_0^\dagger \xi_0).
\label{Lagrangianp}
\end{align}
 In the Appendix \ref{sec:appendix zeromode}, we derive the Hamiltonian corresponding to the Lagrangian density of  Eq.(\ref{Lagrangianp}). 
The Lagrangian in Eq.(\ref{Lagrangianp}) has the form of a constrained system. We identify the constraints  
and impose the additional gauge fixing-like conditions.  Including all of them, they form  
second class constraints.
In Table \ref{constraints}, we show  all the constraints
and gauge fixing-like conditions.
\begin{table}[htbp]
\centering
\begin{tabular}{|c|c|c|c|c|c|c|c|c|} \hline
constraints & $\phi^1(x)$ & $\phi^2(x)$ & $\phi^3$ & $\phi^4$  & $ \phi^5$ & $\phi^6$  & $\phi^7$ &  $\phi^8 $ \\  \hline
                & $\pi_\eta-i\eta^\dagger$ & $\pi_{\eta^\dagger} $ & $\eta_0$ & $\eta_0^\dagger$  & $\pi_{\xi_0},$ & $\pi_{\xi_0^\dagger}$ & $\xi_0$ & $\xi_0^\dagger$ \\ \hline
\end{tabular}
\caption{constraints and gauge-fixing like conditions $\phi^A=0$ $(A=1 \sim 8).$
See Appendix \ref{sec:appendix zeromode}
for the derivation of
the constraints.
}
\label{constraints}
\end{table}

They are used to compute the Dirac bracket \cite{Dirac1964}  among the dynamical variables.  Then we can quantize the field by setting the anti-commutator among the fields based on the Dirac bracket.  The derivation of the Dirac bracket is given in Appendix \ref{sec:appendix zeromode}.  Here we focus on the anti-commutation relations among $\eta$ and $\eta^\dagger$, Eqs.(\ref{acrelations3}-\ref{acrelations4}),
\begin{align}
\{\eta(\bx,t),\eta^{\dagger}(\by,t)\} &= \delta^{(3)}({\bf x}-{\bf y})-\frac{1}{V}, \label{acrelations1}\\
\{\eta(\bx,t),\eta(\by,t)\} &= \{\eta^{\dagger}(\bx,t),\eta^{\dagger}(\by,t)\} = 0.\label{acrelations2}
\end{align} 
In Eqs.(\ref{acrelations1}-\ref{acrelations2}), the field operators $\eta$ and $\eta^\dagger$ satisfy an anti-commutation relation excluding zero mode because $\{\eta_0, \eta_0^\dagger\}
=\{\eta_0, \eta_0\}=\{\eta_0^\dagger, \eta_0^\dagger \}=0$.
As a result, the Hamiltonian can be expanded in terms of the fields $\eta(\bx, t), \eta^\dagger(\bx,t)$ without zero mode as, 
\begin{align}
H= \int d^3\bx \left[\eta^{\dagger}i \boldsymbol{\sigma} \cdot \boldsymbol{\nabla}\eta +\frac{m}{2}(- \eta^{\dagger}i\sigma_2\eta^\dagger + \eta i\sigma_2\eta)\right]
\label{Hamiltonianwozero},
\end{align}
where the last term of Eq.(\ref{H0density}) in Hamiltonian density is dropped because it is proportional to the constraints $\phi^i (i=3,4,7,8)$.
The Hamiltonian is expanded by creation and annihilation operators with non-zero momentum.
From Eq.(A16) in \cite{SalimAdam:2021suq}, the two component chiral field $\eta$, with the zero mode excluded, can be expressed using creation and annihilation operators as,
\begin{align}
\eta(\bx,t) = \int_{\bp \in A} \frac{d^3\bp}{(2\pi)^3\sqrt{2|\bp|}}&\{[a(\bp,t)\phi_{-}(\bn_\bp) - b^\dagger(-\bp,t)\phi_{-}(-\bn_\bp)]e^{i\bp \cdot \bx}  \nn \\ 
&+ [a(-\bp,t)\phi_{-}(-\bn_\bp) - b^\dagger(\bp,t)\phi_{-}(\bn_\bp)]e^{-i\bp \cdot \bx}  \},
\label{expansion}
\end{align}
where $\bn_\bp=\frac{\bp}{|\bp|}$ and the momentum region A is a hemisphere region 
\cite{Adam:2021vbl} defined by,
\bea
A=\{ \bp=|\bp| \bn_\bp, \bn_\bp=\begin{pmatrix} \cos\phi \sin \theta & \sin \phi  \sin \theta & \cos \theta \end{pmatrix} ;  0 \le \phi, \theta < \pi, \bp \ne {\bf 0}\}.
\label{eq:A}
\eea
The north pole which we define to be $\theta=\phi=0$ is included in the momentum region A, while the south pole $\theta=\phi=\pi $ is not.
 Also, the two component spinors $\phi_{\pm}(\pm\bn_\bp)$ are written by the polar angle $\theta$ and the azimuthal angle $\phi$ specifying $\bn_\bp$, 
\begin{align}
&\phi_{+}(\bn_\bp) = \begin{pmatrix} e^{-i\frac{\phi}{2}} \cos \frac{\theta}{2} \\
e^{i\frac{\phi}{2}} \sin \frac{\theta}{2} \end{pmatrix},\quad
\phi_{-}(\bn_\bp) = \begin{pmatrix} -e^{-i\frac{\phi}{2}} \sin \frac{\theta}{2} \\
e^{i\frac{\phi}{2}} \cos \frac{\theta}{2} \end{pmatrix},  
\end{align}
\begin{align}
&\phi_{+}(-\bn_\bp) = i\phi_{-}(\bn_\bp), \quad \phi_{-}(-\bn_\bp) = i\phi_{+}(\bn_\bp).   
\end{align}
In the expansion in Eq.(\ref{expansion}), the two component spinor for $b^+(\bp)$ is $\phi_-(n_\bp)$ so that CP conjugate of the neutrino i.e., anti-neutrino's helicity is right-handed in the massless limit. 
We assume the expansion is valid for the non-zero Majorana mass case so that one can take the smooth massless limit. 
Then the Hamiltonian of single Majorana field, excluding zero mode contribution, is equal to the sum of the non-zero mode contribution of Eq.(A25) in \cite{SalimAdam:2021suq}
as denoted by $h(\bp,t)$ below. Then 
 we start with the following Hamiltonian, 
\begin{align}
    H &=  \int_{\textbf{p}\in A}\frac{d^3\bp}{(2\pi)^32|\textbf{p}|}|\textbf{p}|[a^{\dagger}(\textbf{p},t)a(\textbf{p},t) + b^{\dagger}(\textbf{p},t)b(\textbf{p},t) \nn \\
&\qquad \qquad \qquad \qquad +a^{\dagger}(-\textbf{p},t)a(-\textbf{p},t) + b^{\dagger}(-\textbf{p},t)b(-\textbf{p},t) ] \nn \\
    &+ m\int_{\textbf{p}\in A}\frac{d^3\bp}{(2\pi)^32|\textbf{p}|}[-ia(\textbf{p},t)a(-\textbf{p},t) -i b(\textbf{p},t)b(-\textbf{p},t) + h.c.], \notag\\
 &= \sum_{\textbf{p}\in A}h(\textbf{p},t).
\label{eq:Hamiltonian}
\end{align}
In the Hamiltonian,  $h(\textbf{p},t)$  is written in terms of the set of operators $\{a(\pm \bp,t)$, $
b(\pm
\bp,t)$, $a^\dagger(\pm \bp,t)$, $b^\dagger(\pm \bp,t) \}$ where $\bp$ is a  momentum in A region in Eq.(\ref{eq:A}). Furthermore dimensionless operators $\alpha(\bp,t)$ and $\beta(\bp,t)$ are introduced,
\begin{align}
    \alpha(\textbf{p},t) = \frac{a(\textbf{p},t)}{\sqrt{(2\pi)^3 2|\textbf{p}|\delta^{(3)}(0)}}, \quad
 \beta(\textbf{p},t) = \frac{b(\textbf{p},t)}{\sqrt{(2\pi)^3 2|\textbf{p}|\delta^{(3)}(0)}}.
\label{11}
\end{align}
The creation operators and annihilation operators satisfy the following anti-commutation
relations,
\bea
 \{ \alpha(\bp, t), \alpha^\dagger(\bq, t) \}= \{ \beta(\bp,t), \beta^\dagger(\bq,t) \}=\delta_{\bp \bq}.
\eea
By using them,  $h(\bp,t)$ 
\footnote{ One can show $h(\textbf{p},t)$ is independent of time. Then $h(\textbf{p},t)=h(\textbf{p},t_i)$.}
 is written as, 
\begin{align}
    h(\bp,t) &= |\bp|[N_{\alpha}(\bp,t)+N_{\beta}(\bp,t)+N_{\alpha}(-\bp,t)+N_{\beta}(-\bp,t)] \nn \\
    &-im[B_{\alpha}(\bp,t)+B_{\beta}(\bp,t) - B^{\dagger}_{\alpha}(\bp,t)-B^{\dagger}_{\beta}(\bp,t)] \label{16}.
\end{align}
In Eq.(\ref{16}), the following bilinear operators are introduced,
\begin{align}
B_{\alpha}(\textbf{p}, t) &= \alpha(\textbf{p},t)\alpha(-\textbf{p},t), \label{Coopera}\\ 
B_{\beta}(\textbf{p},t) &= \beta(\textbf{p},t)\beta(-\textbf{p},t), \label{Cooperb}\\
N_{\alpha}(\pm \textbf{p},t) &= \alpha^{\dagger}(\pm\textbf{p},t)\alpha(\pm \textbf{p},t), \label{Na}\\
N_{\beta}(\pm \textbf{p},t) &= \beta ^{\dagger}(\pm \textbf{p},t)
\beta(\pm \textbf{p},t).\label{Nb}
\end{align}
Througout this paper, we call the bilinear operators  $B_{\alpha}(\textbf{p},t)$ ($B_{\beta}(\textbf{p},t)$) as
Cooper pair operator since this operator annihilates a pair of the neutrinos (anti-neutrinos)  with opposite momentum.   
$B_{\alpha}(\textbf{p},t)$ and $N_{\alpha}(\textbf{p},t)$ satisfy the commutation relations,
\begin{align}
    [N_{\alpha}(\pm\bp,t),B_{\alpha}(\bq,t)] &= -B_{\alpha}(\textbf{p},t)\delta_{\textbf{pq}}, \label{12}\\
    [N_{\alpha}(\pm\bp,t),B^{\dagger}_{\alpha}(\bq,t)] &= B^{\dagger}_{\alpha}(\textbf{p},t)\delta_{\textbf{pq}},\label{13}\\
    [B_{\alpha}(\bp,t),B^{\dagger}_{\alpha}(\bq,t)] &= (1-N_{\alpha}(\textbf{p},t)-N_{\alpha}(-\textbf{p},t))\delta_{\textbf{pq}}, \label{14}\\
    [N_{\alpha}(\bp,t),N_{\alpha}(\bq,t)] &=0. \label{15}
\end{align}
The bilinear operators for anti-neutrinos $B_{\beta}(\textbf{q},t)$ and $N_{\beta}(\textbf{q},t)$ satisfy the same commutation 
relations as in Eq.(\ref{12}-\ref{15}).
Using the commutation relations, the Hamiltonians for different momentum, $\textbf{p} \ne \textbf{q}$ commute each other,   
\begin{align}
    [h(\textbf{p},t),h(\textbf{q},t)] = 0. \label{17}
\end{align}
 Hereafter the set of the operators $\{\alpha(\pm \bp,t), \beta(\pm
\bp,t), \alpha^\dagger(\pm \bp,t), \beta^\dagger(\pm \bp,t) \}$ and their bilinear operators in Eqs.(\ref{Coopera}-\ref{Nb}) which appear in $h(\bp,t)$ are called as  operators of $\bp$ sectors. For instance, the operator $\alpha(\bp) $ and $\alpha(-\bp)$ with $\bp \in A$ are classified as the operators in the same $\bp$ sector.
\section{Time evolution of operators and Bogoliubov transformation }\label{sec: Bogoliubov transformation} 
In this section, we first show the time evolution of the creation and annihilation operators using the Hamiltonian
in Eq.(\ref{eq:Hamiltonian}).  We also derive the time evolution of the
 Cooper pair operators.  Next, the relation between the eigenstates of the lepton number operator defined at production time $t_i$ and detection time $t_f$ is  written with the Bogoliubov transformation.
\subsection{Time evolution of operators}
The time evolution of the annihilation operators are given by,
\begin{align}
\alpha(\textbf{p},t_f) = e^{iH\tau} \alpha(\textbf{p},t_i)e^{-iH\tau},\quad
\beta(\textbf{p},t_f) = e^{iH\tau} \beta(\textbf{p},t_i)e^{-iH\tau},
\label{7}
\end{align}
where $\tau=t_f-t_i$.  
We define the vacuum $\ket{0, t}$ as, 
\bea
\alpha(\textbf{p},t) \ket{0, t}=\beta(\textbf{p},t)\ket{0, t}=0. 
\label{vacuum}
\eea
$\ket{0, t}$ is the eigenstate of the lepton number where the lepton number operator is defined by,
\bea
&&L(t)=\sum_{\textbf{p}\in A} L(\textbf{p},t), \nn \\
&&L(\textbf{p},t) = N_{\alpha}(\textbf{p},t)-N_{\beta}(\textbf{p},t)+N_{\alpha}(-\textbf{p},t)-N_{\beta}(-\textbf{p},t).
\label{LeptonNumber}
\eea
From Eqs.(\ref{vacuum}-\ref{LeptonNumber}),
$\ket{0, t}$ is the state with the zero eigenvalue of $L(t)$.
Since the following relation holds true,
\begin{align}
    \alpha(\textbf{p},t_f)|0,t_f\rangle =e^{iH\tau}  \alpha(\textbf{p},t_i)e^{-iH\tau}|0,t_f\rangle = 0, \\
\beta(\textbf{p},t_f)|0,t_f\rangle =e^{iH\tau}  \beta(\textbf{p},t_i)e^{-iH\tau}|0,t_f\rangle = 0,
 \label{8}
\end{align}
the two vacua $|0,t_f\rangle$ and $|0,t_i\rangle$ are related to each other as,
\begin{align}
    |0,t_f\rangle = e^{iH\tau} |0,t_i\rangle. \label{9}
\end{align}
In the present formulation, the vacuum depends on time as can be seen from Eq.(\ref{9}).
To construct the Fock states, it is convenient to introduce the $\bp$ sector vacuum $\ket{0,t}_{\textbf{p}}$.
 The vacuum is expressed by the direct product of the  $\textbf{p}$ sector vacuum,
\begin{align}
    \ket{0,t} = \prod_{\textbf{p}\in A}\ket{0,t}_{\textbf{p}}. \label{prodstate}
\end{align}
On this $\bp$ sector vacuum, operators of $\bp$ sector act.
 Using the property Eq.(\ref{17}) and the definition Eq.(\ref{prodstate}), we rewrite the time evolution for the vacuum in Eq.(\ref{9}) as,
\begin{align}
    |0,t_f\rangle = \prod_{\textbf{p}\in A}  |0,t_f\rangle_{\textbf{p}}=   e^{iH\tau} |0,t_i\rangle  =\prod_{\textbf{p}\in A}e^{ih(\textbf{p},t_i)\tau} |0,t_i\rangle_{\textbf{p}}. \label{18}
\end{align}
Then one can show that the time evolution of the vacuum of $\bp$ sector is given as,
\bea
|0,t_f\rangle_{\textbf{p}}&=&e^{ih(\bp,t_i)\tau} |0,t_i\rangle_{\textbf{p}}, \\
    &=& \sum_{n}\frac{1}{n!} (\tau ih(\bp,t_i))^n|0,t_i\rangle_{\textbf{p}}. \label{19}
\eea
 Next we study the time evolution of the Cooper pair operators.  For this purpose, we show the
 time evolution of $\alpha(\textbf{p},t)$ and $\beta(\textbf{p},t)$ from the production time $t_i$ to the detection time $t_f$ \cite{Adam:2021qiq, SalimAdam:2021suq},
\begin{align}
    \alpha(\pm \textbf{p},t_f) = \left(\cos E_\bp\tau-i\frac{|\textbf{p}|}{E_\bp}\sin E_\bp\tau \right)\alpha(\pm \textbf{p},t_i) \mp \frac{m}{E_\bp}\sin E_\bp\tau \alpha^\dagger(\mp \textbf{p},t_i), \label{atime}\\
    \beta(\pm \textbf{p},t_f) = \left(\cos E_\bp\tau-i\frac{|\textbf{p}|}{E_\bp}\sin E_\bp\tau \right)\beta(\pm \textbf{p},t_i) \mp \frac{m}{E_\bp}\sin E_\bp\tau \beta^\dagger(\mp \textbf{p},t_i),\label{btime}
\end{align}
where $E_\bp=\sqrt{\textbf{p}^2+m^2}$. \\
We can  derive the time evolution of the Cooper pair operator in Eq.(\ref{Coopera}) using Eq.\eqref{atime},
\begin{align}
    B_\alpha(\textbf{p},t_f) &= \alpha(\textbf{p},t_f)\alpha(-\textbf{p},t_f), \notag \\
    &= \left(\cos E_\bp\tau-i\frac{|\textbf{p}|}{E_\bp}\sin E_\bp\tau \right)^2B_\alpha(\textbf{p},t_i) - \left(\frac{m}{E_\bp}\sin E_\bp\tau\right)^2B_\alpha^\dagger(\textbf{p},t_i) \notag \\ 
    &+ \left(\cos E_\bp\tau-i\frac{|\textbf{p}|}{E_\bp}\sin E_\bp\tau \right)\frac{m}{E_\bp}\sin E_\bp\tau \left(1-N_\alpha(\textbf{p}, t_i)-N_\alpha(-\textbf{p}, t_i) \right).
\label{Cooper}
\end{align}
 For the Cooper pair operator for anti-neutrinos $B_\beta(\textbf{p},t_f)$ in Eq.(\ref{Cooperb}), one can derive the relation similar to Eq.(\ref{Cooper}) by replacing 
 $\alpha$ with $\beta$ in Eq.(\ref{Cooper}).
\subsection{Bogoliubov transformation}
We first define the set of the states defined at arbitrary time $t$ by applying the  Cooper pair operators on the vacuum $|0,t \rangle_{\textbf{p}}$,
\begin{align}
   |2,t \rangle_{\textbf{p}} 
   &=\frac{1}{\sqrt{2}} \left[B^{\dagger}_{\alpha}(\textbf{p},t )+B^{\dagger}_{\beta}(\textbf{p},t)\right]|0,t \rangle_{\textbf{p}},
   \label{21}\\
   |4,t \rangle_{\textbf{p}} 
   &= B^{\dagger}_{\alpha}(\textbf{p},t )B^{\dagger}_{\beta}(\textbf{p},t )|0,t \rangle_{\textbf{p}}, \label{22}
\end{align}
where $|2,t \rangle_{\textbf{p}}$ and $ |4,t \rangle_{\textbf{p}} $ imply two particles and four particles 
states respectively.
The $S$ operator $S_{\bp}(\tau)= e^{-i h(\textbf{p})\tau} $ relates the set of the ket vectors $ 
\begin{pmatrix} _{\bp}\langle 0, t_f| & _{\bp}\langle 2,t_f| & _{\bp}\langle 4,t_f| \end{pmatrix} $ and those at $t=t_i$ as,
\begin{align}
\begin{pmatrix} _{\bp}\langle 0, t_f| & _{\bp}\langle 2,t_f| & _{\bp}\langle 4,t_f| \end{pmatrix} =
\begin{pmatrix} _{\bp}\langle 0, t_i| & _{\bp}\langle 2,t_i| & _{\bp}\langle 4,t_i| \end{pmatrix}  e^{-i h(\textbf{p}) \tau},
\end{align}
 As for bra vectors, the time evolution is expressed as, 
\begin{align}
\begin{pmatrix}|0,t_f\rangle_{\textbf{p}} &
    |2,t_f\rangle_{\textbf{p}} &
    |4,t_f\rangle_{\textbf{p}} \end{pmatrix}
&=
   e^{i h(\textbf{p})\tau} \left(\begin{array}{ccc}
    |0,t_i\rangle_{\textbf{p}} &
    |2,t_i\rangle_{\textbf{p}} &
    |4,t_i\rangle_{\textbf{p}}
    \end{array}\right) \nn \\
&=\begin{pmatrix}
    |0,t_i\rangle_{\textbf{p}} &
    |2,t_i\rangle_{\textbf{p}} &
    |4,t_i\rangle_{\textbf{p}}
\end{pmatrix} 
\begin{pmatrix}
G_{11}(\textbf{p},\tau) & G_{12}(\textbf{p},\tau) & G_{13}(\textbf{p},\tau) \\
     G_{21}(\textbf{p},\tau ) & G_{22}(\textbf{p},\tau) & G_{23}(\textbf{p},\tau) \\
     G_{31}(\textbf{p},\tau) & G_{32}(\textbf{p},\tau) & G_{33}(\textbf{p},\tau) 
\end{pmatrix},
 \label{TEGmatrix}
\end{align}
where $G_{ij}(\textbf{p},\tau)$ denotes the matrix elements of the operator $S^\dagger_\bp= e^{i h(\textbf{p})\tau} $ among the states at $t=t_i$,
\begin{align}
\begin{pmatrix}
G_{11}(\textbf{p},\tau) & G_{12}(\textbf{p},\tau) & G_{13}(\textbf{p},\tau) \\
     G_{21}(\textbf{p},\tau ) & G_{22}(\textbf{p},\tau) & G_{23}(\textbf{p},\tau) \\
     G_{31}(\textbf{p},\tau) & G_{32}(\textbf{p},\tau) & G_{33}(\textbf{p},\tau) 
\end{pmatrix}
=\begin{pmatrix} _{\bp}\langle 0, t_i| \\ _{\bp}\langle 2,t_i| \\ _{\bp}\langle 4,t_i| \end{pmatrix} 
 e^{i h(\textbf{p})\tau} \begin{pmatrix}
    |0,t_i\rangle_{\textbf{p}} &
    |2,t_i\rangle_{\textbf{p}} &
    |4,t_i\rangle_{\textbf{p}}
    \end{pmatrix}.
\label{Gmatrix}
\end{align}
The matrix elements of $G(\bp,\tau)$ and the matrix elements of $S^\dagger_\bp$ have the following correspondance,
\begin{align}
& _{\textbf{p}}\langle 0,t_i |0,t_f\rangle_{\textbf{p}}=\, _{\bp}\langle 0,t_i |S^\dagger_\bp|0,t_i\rangle_{\textbf{p}}=G_{11}(\textbf{p},\tau), \\
& _{\textbf{p}}\langle 2,t_i |0,t_f\rangle_{\textbf{p}}=\, _{\bp}\langle 2,t_i |S^\dagger_\bp|0,t_i\rangle_{\textbf{p}}=G_{21}(\textbf{p},\tau),\\
& _{\textbf{p}}\langle 4,t_i |0,t_f\rangle_{\textbf{p}}=\, _{\bp}\langle 4,t_i |S^\dagger_\bp|0,t_i\rangle_{\textbf{p}}=G_{31}(\textbf{p},\tau),\\
& _{\textbf{p}}\langle 0,t_i |2,t_f\rangle_{\textbf{p}}=\, _{\bp}\langle 0,t_i |S^\dagger_\bp|2,t_i\rangle_{\textbf{p}}=G_{12}(\textbf{p},\tau),\\
& _{\textbf{p}}\langle 2,t_i |2,t_f\rangle_{\textbf{p}}=\, _{\bp}\langle 2,t_i |S^\dagger_\bp|2,t_i\rangle_{\textbf{p}}=G_{22}(\textbf{p},\tau),\\
& _{\textbf{p}}\langle 4,t_i |2,t_f\rangle_{\textbf{p}}=\, _{\bp}\langle 4,t_i |S^\dagger_\bp|2,t_i\rangle_{\textbf{p}}=G_{32}(\textbf{p},\tau),\\
& _{\textbf{p}}\langle 0,t_i |4,t_f\rangle_{\textbf{p}}=\, _{\bp}\langle 0,t_i |S^\dagger_\bp|4,t_i\rangle_{\textbf{p}}=G_{13}(\textbf{p},\tau),\\
& _{\textbf{p}}\langle 2,t_i |4,t_f\rangle_{\textbf{p}}=\, _{\bp}\langle 2,t_i |S^\dagger_\bp|4,t_i\rangle_{\textbf{p}}=G_{23}(\textbf{p},\tau),\\
& _{\textbf{p}}\langle 4,t_i |4,t_f\rangle_{\textbf{p}}=\, _{\bp}\langle 4,t_i |S^\dagger_\bp|4,t_i\rangle_{\textbf{p}}=G_{33}(\textbf{p},\tau).
\end{align}
In the following, we present how one can derive the elements of the matrix $G(\bp,\tau)$. 
We consider the time evolution of each eigenstates $|0,t_i\rangle_{\textbf{p}}$, $|2,t_i\rangle_{\textbf{p}}$ and $|4,t_i\rangle_{\textbf{p}}$ using Eq.(\ref{19}) and Eq.(\ref{TEGmatrix}). 
Since one can expand the unitary operator $e^{i h(\textbf{p})\tau} $ as a series
 $\sum_{n}\frac{1}{n!} (\tau ih(\textbf{p}))^n|0,t_i\rangle_{\textbf{p}} $, 
one can study the action of the operator $ \tau ih(\textbf{p})$ on the three 
representative states as,
\begin{align}
    (\tau ih(\textbf{p}))\left(\begin{array}{ccc}
    |0,t_i\rangle_{\textbf{p}} &
    |2,t_i\rangle_{\textbf{p}} &
    |4,t_i\rangle_{\textbf{p}}
    \end{array}\right)
    &=i\sqrt{2}m\tau \left(\begin{array}{ccc}
    |0,t_i\rangle_{\textbf{p}} &
    |2,t_i\rangle_{\textbf{p}} &
    |4,t_i\rangle_{\textbf{p}}
    \end{array}\right)
    \left(\begin{array}{ccc}
     0 & -i & 0 \\
     i & \sqrt{2}k & -i \\
     0 & i & 2\sqrt{2}k
    \end{array}\right), \notag \\
    &=\left(\begin{array}{ccc}
    |0,t_i\rangle_{\textbf{p}} &
    |2,t_i\rangle_{\textbf{p}} &
    |4,t_i\rangle_{\textbf{p}}
    \end{array}\right)\A(i\sqrt{2}m\tau),
 \label{Amatrix}
\end{align}
where  
\begin{align}
\A= \left(\begin{array}{ccc}
     0 & -i & 0 \\
     i & \sqrt{2}k & -i \\
     0 & i & 2\sqrt{2}k
    \end{array}\right),
\label{Amatrix2}
\end{align}
and $k=\frac{|\textbf{p}|}{m}$.
Therefore the action of $e^{i h(\textbf{p})\tau}$ is given by,
\begin{align}
   e^{i h(\textbf{p})\tau}\left(\begin{array}{ccc}
    |0,t_i\rangle_{\textbf{p}} &
    |2,t_i\rangle_{\textbf{p}} &
    |4,t_i\rangle_{\textbf{p}}
    \end{array}\right)
    &=\left(\begin{array}{ccc}
    |0,t_i\rangle_{\textbf{p}} &
    |2,t_i\rangle_{\textbf{p}} &
    |4,t_i\rangle_{\textbf{p}}
    \end{array}\right)e^{ \A (i\sqrt{2}m\tau)}
\label{Anmatrix},
\end{align}
From Eq.(\ref{Gmatrix}), one finds, 
\begin{align}
G(\textbf{p},\tau) =e^{ \A (i\sqrt{2}m\tau)}.
\label{G}
\end{align}
The matrix form of $e^{ \A (i\sqrt{2}m\tau)}$ can be obtained 
by diagonalizing matrix $\A$ using a unitary matrix V. From Eq.(\ref{Amatrix2}),
\begin{align}
    \omega_i\delta_{ij}=V_{ik} \A_{kl}V^{-1}_{lj},
\end{align}
where $\omega_i(i=1\sim 3)$ are the eigenvalues of $\tilde{A}$.
Then the matrix $\A$ is written as,
\begin{align}
&\A=V^{-1}\Omega V, 
\label{AV}
\end{align}
where $\Omega$ is a real diagonal matrix of the eigenvalues of $\A$,
\begin{align}
\Omega = \begin{pmatrix}
\omega_1 & 0 &0 \\
    0 & \omega_2 & 0 \\
    0 & 0 & \omega_3
\end{pmatrix}.
\end{align}
Then the matrix $G$ in Eq.(\ref{G}) is also written as,
\begin{align}
e^{ \A (i\sqrt{2}m\tau)}= V^{-1} e^{ \Omega (i\sqrt{2}m\tau)} V
=V^{-1} \begin{pmatrix}e^{i \omega_1 \sqrt{2}m\tau} & 0 & \\
0 & e^{i \omega_2 \sqrt{2}m\tau} & 0 \\
0 & 0 & e^{i \omega_3 \sqrt{2}m\tau} \end{pmatrix} V.
\label{ExpA}
\end{align}
From Eq.(\ref{AV}), $V^T$ satisfies the following equation,
\begin{align}
& \A^T V^T=V^T \Omega.
\end{align}
The rest of the task is to find the eigenvalues of $\A^T$ and the matrix $V^T$.
$V^T$ consists of the three eigenvectors for $\A^T$. 
The eigenvalues of $\A^T$ can be obtained via,
\begin{align}
    &\det(\A^{T}-\omega I)=
    \det\left(\begin{array}{ccc}
     -\omega & i & 0 \\
     -i & \sqrt{2}k -\omega & i \\
     0 & -i & 2\sqrt{2}k -\omega
    \end{array}\right)=0,\notag 
\end{align}
and they are given by,
\begin{align}
    &\omega_1=\sqrt{2}k_{+}, \ \omega_2= \sqrt{2} k, \ \omega_3=\sqrt{2}{k_-},
\end{align}
where 
$k_{\pm}= k\pm\sqrt{k^2+1}$.
With the eigenvalues obtained,  we will  find the eigenvectors.
We write $V^T $ with three complex vectors ${\bf v}_i (i =1 \sim 3)$ as,
\begin{align}
V^T=\begin{pmatrix} {\bf v}^T_1 & {\bf v}^T_2 &  {\bf v}^T_3 \end{pmatrix}.
\end{align}
For $\omega_2=\sqrt{2}k$,  the corresponding eigenvector is,
\begin{align}
  {\bf v}_2^T  =\frac{1}{\sqrt{2k^2+2}}\left(\begin{array}{c}
         1  \\
         -\sqrt{2}ki  \\
         1
    \end{array}\right).
\end{align}
For $\omega_1=\sqrt{2}k_+$ and  $\omega_3=\sqrt{2}k_-$, the corresponding eigenvector is  given respectively as,
\begin{align}
     {\bf v}_1^T 
    =\frac{1}{\sqrt{2k^2+2}}\left(\begin{array}{c}
         \frac{\sqrt{2}k_-}{2}  \\
         i  \\
         \frac{\sqrt{2}k_+}{2}
    \end{array}\right),\quad
{\bf v}_3^T 
    =\frac{1}{\sqrt{2k^2+2}}\left(\begin{array}{c}
         \frac{\sqrt{2}k_+}{2}  \\
         i  \\
         \frac{\sqrt{2}k_-}{2}
    \end{array}\right). \quad
\end{align}
Therefore, the unitary matrix $V$ and $V^{-1}$ are given by,
\begin{align}
    V=\frac{1}{\sqrt{2k^2+2}}\left(\begin{array}{ccc}
     \frac{\sqrt{2}k_-}{2} & i & \frac{\sqrt{2}k_+}{2} \\
     1 & -\sqrt{2}ki & 1 \\
     \frac{\sqrt{2}k_+}{2} & i & \frac{\sqrt{2}k_-}{2}
    \end{array}\right),
\label{V}
\quad
    V^{-1}=V^{\dagger}=\frac{1}{\sqrt{2k^2+2}}\left(\begin{array}{ccc}
     \frac{\sqrt{2}k_-}{2} & 1 & \frac{\sqrt{2}k_+}{2} \\
     -i & \sqrt{2}ki & -i \\
     \frac{\sqrt{2} k_+}{2} & 1 & \frac{\sqrt{2}k_-}{2}
    \end{array}\right).
\end{align}
By substituting the eigenvalues to Eq.(\ref{ExpA}), from Eq.(\ref{V})
the matrix $G(\bp,\tau)$ is given by,
\begin{align}
    G(\textbf{p},\tau) 
&=V^{\dagger}\left(\begin{array}{ccc}
     e^{2ik_{+}m\tau} & 0 & 0 \\
     0 & e^{2ikm\tau} & 0 \\
     0 & 0 & e^{2ik_{-}m\tau}
    \end{array}\right)V, \nn \\
&= e^{2i|\bp|\tau}
\begin{pmatrix}
f(\bp,\tau)^2 & \sqrt{2} f(\bp,\tau)g(\bp,\tau) & g(\bp,\tau)^2 \\  
-\sqrt{2}f(\bp,\tau)g(\bp,\tau) & 1-2g(\bp,\tau)^2 & \sqrt{2}f^*(\bp,\tau)g(\bp,\tau) \\ 
g(\bp,\tau)^2 & -\sqrt{2} f^\ast(\bp,\tau) g(\bp,\tau) & f^*(\bp,\tau)^2
\end{pmatrix}, \label{Gm}
\end{align}
where the functions $f(\textbf{p},\tau)$ and $g(\textbf{p},\tau)$ are given as,
\bea
f(\textbf{p},\tau)&=& \cos E_\textbf{p}\tau
-i\frac{|\textbf{p}|}{E_\textbf{p}}\sin E_\textbf{p}\tau, \label{eq:f}\\
 g(\textbf{p},\tau)&=&\frac{m}{E_\textbf{p}}\sin E_\textbf{p}\tau. \label{eq:g}
\eea
In addition, the functions $f(\bp,\tau)$ and $g(\bp,\tau)$ satisfy the following relation,
\bea
\left|f(\bp,\tau)\right|^2+\left|g(\bp,\tau)\right|^2=1. \label{unitarityfg}
\eea
Thus, from Eq.(\ref{TEGmatrix}), we can rewrite the vacuum $|0,t_f \rangle_{\bp}$ as,
\begin{align}
     |0,t_f\rangle_{\textbf{p}} 
   &=G_{11}(\textbf{p},\tau)|0,t_i\rangle_{\textbf{p}} + G_{21}(\textbf{p},\tau)|2,t_i\rangle_{\textbf{p}} + G_{31}(\textbf{p},\tau)|4,t_i\rangle_{\textbf{p}}, \notag \\
     &=G_{11}(\textbf{p},\tau)
\exp \left[\frac{G_{21}(\textbf{p},\tau)}{\sqrt{2}G_{11}(\textbf{p},\tau)}\left(B^{\dagger}_{\alpha}(\textbf{p},t_i) + B^{\dagger}_{\beta}(\textbf{p},t_i)\right)\right]|0,t_i\rangle_{\textbf{p}} \label{0tevo}.
\end{align}
In the second line of Eq.(\ref{0tevo}), we write the relation between $\ket{0,t_f}_\bp$  and  $\ket{0,t_i}_\bp$ with the Bogoliubov transformation.
The other states, $|2,t_f \rangle_{\textbf{p}}$ and  $|4,t_f \rangle_{\textbf{p}}$ are also expressed by the superposition of the states $\ket{0, t_i}_\bp$, $\ket{2,t_i}_\bp$  and $\ket{4,t_i}_\bp$ with Eq.(\ref{TEGmatrix}) and Eq.(\ref{Gm}).
These relations are also expressed with the Bogoliubov transformation similar to Eq.(\ref{0tevo}).

\section{Probability}\label{sec:Probability}
In this section,  we compute the time-dependent transition probability for the neutrino with 
momentum $\bp$.
Since the Fock state is given by the direct product of the state specified by each momentum sector in Eq.(\ref{eq:A}), 
the transition amplitude is also given by the product of the amplitude in each momentum sector.
To obtain the transition amplitude of a single neutrino with momentum $\pm \bp \ (\bp\in A)$, it is not sufficient to know the transition amplitude of the state
$\alpha^\dagger(\pm\bp, t_i) \ket{0, t_i}_{\bp}$. 
As shown below, one needs to specify the states for  all the momentum sectors  specified by Eq.(\ref{eq:A}). Then
the one-particle state with momentum $\pm \bp$ is expressed as,
\bea
\alpha^\dagger(\pm \bp,t_i) \ket{0,t_i}=  \alpha^\dagger(\pm \bp,t_i) \ket{0, t_i}_{\bp} \prod_{\bq\ne \bp} \ket{0, t_i}_{\bq}, \quad
\bp, \bq \in A,
\label{state} 
\eea
where we divide the momentum sectors into $\bp$ sector with one-particle  and
the other $\bq$ sectors where  particles are absent.
Then we 
calculate the transition amplitudes for both sectors separately. 
For $\bp$ sector, one obtains the $S$ matrix elements for 
the neutrino transitions from  initial one-particle state to the possible
final states by defining the $S$-matrix operator called as $\mathcal{S}_\bp$.
In the other $\bq$ sector, the vacuum $|0, t_i\rangle_{\bq}$  transits to  the states of even number of particles including two or four particle states. We define the S-matrix operator $S_\bq$
representing the transitions. Finally, we formulate the time-dependent oscillation probability with the matrix elements of  $\mathcal{S}_\bp$ and  $S_\bq$.
\begin{figure}[htbp]
\begin{center}
\includegraphics[width=0.62\textwidth]{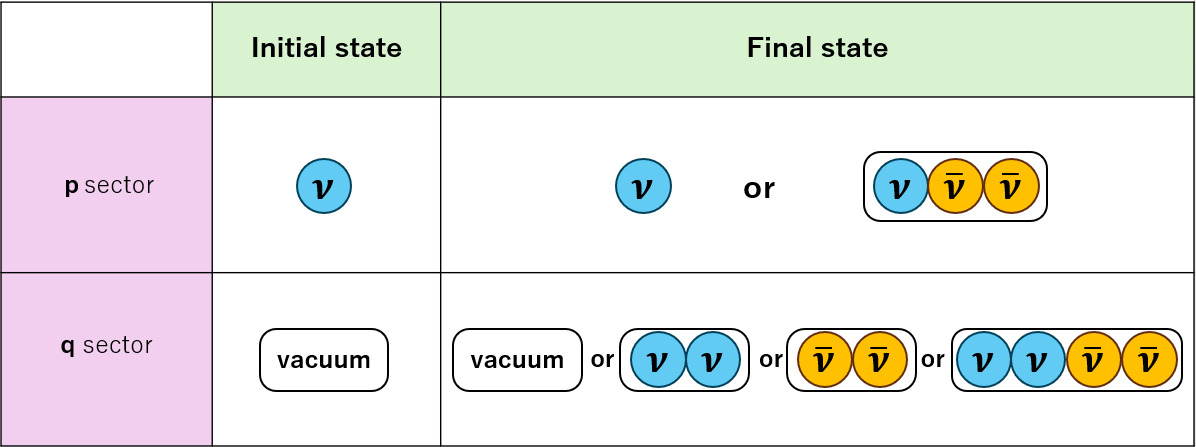}
   \caption{Initial states and all possible final states of the transitions in each momentum sector}
\label{each sectors}
\end{center}
Fig.\ref{each sectors} summarizes the transitions from the initial state with lepton number $+1$ and
the state with lepton number
 $0$. In the $\bp$ sector, the initial state is a one-particle state, and in the $\bq$ sector, the initial state is a vacuum state. The possible final states from each initial state are also shown in the right side of the Fig.\ref{each sectors}.
\end{figure}

\subsection{S-Matrix}
In the momentum sector $\bp$, we  consider the $\mathcal{S}_\bp$ operator representing neutrino oscillation among the states with the lepton number $\pm 1$, based on Eqs.(\ref{atime}),(\ref{Cooper}) and (\ref{0tevo}). 
The matrix elements  for $\mathcal{S}_\bp$  operator are  defined as
\bea
&&\mathcal{S}_{\bp}^{11}= \, _\bp\langle 0,t_i| \alpha(\textbf{p},t_i)\mathcal{S}_{\bp} \alpha^{\dagger}(\textbf{p},t_i)\ket{0,t_i}_{\bp}, \\
&&\mathcal{S}_{\bp}^{13}=  \, _\bp\langle 0,t_i| \alpha(\textbf{p},t_i)\mathcal{S}_{\bp} B_{\beta}^{\dagger}(\textbf{p},t_i)\alpha^{\dagger}(\textbf{p},t_i) \ket{0,t_i}_{\bp}, \\
&&\mathcal{S}_{\bp}^{31}=  \, _\bp\langle 0,t_i| B_{\beta}(\textbf{p},t_i)\alpha(\textbf{p},t_i) \mathcal{S}_{\bp} \alpha^{\dagger}(\textbf{p},t_i)\ket{0,t_i}_{\bp}, \\
&&\mathcal{S}_{\bp}^{33}=  \, _\bp\langle 0,t_i| B_{\beta}(\textbf{p},t_i)\alpha(\textbf{p},t_i)\mathcal{S}_{\bp} B_{\beta}^{\dagger}(\textbf{p},t_i)\alpha^{\dagger}(\textbf{p},t_i)\ket{0,t_i}_{\bp}.
\eea
Using the matrix elements for $\mathcal{S}_\bp$, the relations among the states defined at $t=t_f$ and $t=t_i$ are given by,
\begin{align}
\begin{pmatrix}
\alpha^{\dagger}(\textbf{p},t_f)\ket{0,t_f}_{\bp} \\
B_{\beta}^{\dagger}(\textbf{p},t_f)\alpha^{\dagger}(\textbf{p},t_f)\ket{0,t_f}_{\bp}
\end{pmatrix} &= 
\begin{pmatrix}
\mathcal{S}_{\bp}^{11 \ast } & \mathcal{S}_{\bp}^{13 \ast} \\
\mathcal{S}_{\bp}^{31 \ast} & \mathcal{S}_{\bp}^{33 \ast}
\end{pmatrix}
\begin{pmatrix}
\alpha^{\dagger}(\textbf{p},t_i)\ket{0,t_i}_{\bp} \\
B_{\beta}^{\dagger}(\textbf{p},t_i)\alpha^{\dagger}(\textbf{p},t_i)\ket{0,t_i}_{\bp}
\end{pmatrix} \label{Spmatrix}.
\end{align}
 The state with lepton number $+1$ at $t=t_f$  is given by the following superposition of the states defined at $t=t_i$,
\begin{align}
    \alpha^{\dagger}(\textbf{p},t_f)\ket{0,t_f}_{\bp} &= 
    \alpha^{\dagger}(\textbf{p},t_f)G_{11}(\textbf{p},\tau)\exp \left[\frac{G_{21}(\textbf{p},\tau)}{\sqrt{2}G_{11}(\textbf{p},\tau)}\left(B^{\dagger}_{\alpha}(\textbf{p},t_i) + B^{\dagger}_{\beta}(\textbf{p},t_i)\right)\right]|0,t_i\rangle_{\textbf{p}},  \notag \\ 
    &=  e^{2i|\textbf{p}|\tau}f(\textbf{p},\tau)\alpha^{\dagger}(\textbf{p},t_i) \ket{0,t_i}_{\bp}
    +e^{2i|\textbf{p}|\tau}(-g(\textbf{p},\tau))B_{\beta}^{\dagger}(\textbf{p},t_i)\alpha^{\dagger}(\textbf{p},t_i)\ket{0,t_i}_{\bp} \label{finalstate+1},
\end{align}
 where in the first line of the equation above, we use Eq.\eqref{atime} and Eq.(\ref{0tevo}). 
Similarly, for the three-particle state with lepton number $-1$ is written with the superposition of the states at $t=t_i$ with the lepton number $\mp1$,
\begin{align}
    B_{\beta}^{\dagger}(\textbf{p},t_f)\alpha^{\dagger}(\textbf{p},t_f)\ket{0,t_f}_{\bp} 
    &=  e^{2i|\textbf{p}|\tau}g(\textbf{p},\tau)\alpha^{\dagger}(\textbf{p},t_i) \ket{0,t_i}_{\bp}
    +e^{2i|\textbf{p}|\tau} f^\ast(\textbf{p},\tau) B_{\beta}^{\dagger}(\textbf{p},t_i)\alpha^{\dagger}(\textbf{p},t_i)\ket{0,t_i}_{\bp}, \label{finalstate-1}
\end{align}
where anti-particle version of Eq.(\ref{Cooper}) and Eqs.(\ref{13}-\ref{14}) are used.
From Eqs.(\ref{finalstate+1}-\ref{finalstate-1}), the matrix elements for $\mathcal{S}_\bp $ operator in Eq.(\ref{Spmatrix}) are given by,
\begin{align}
\begin{pmatrix}
\mathcal{S}_{\bp}^{11 *} & \mathcal{S}_{\bp}^{13 *} \\
\mathcal{S}_{\bp}^{31 *} & \mathcal{S}_{\bp}^{33 *}
\end{pmatrix}
=
 e^{2i|\textbf{p}|\tau} \begin{pmatrix}
f(\textbf{p},\tau) & -g(\textbf{p},\tau) \\
g(\textbf{p},\tau) & f^*(\textbf{p},\tau)
\end{pmatrix}\label{Spcomponent},
\end{align}
where $f(\textbf{p},\tau)$ and $g(\textbf{p},\tau)$ are given in Eq.(\ref{eq:f}) and Eq.(\ref{eq:g}) respectively. 
The matrix elements for $\mathcal{S}_\bp$ operator for the transition 
among the anti-neutrino state and the state with a neutrino Cooper pair plus an anti-neutrino
are the same as those in  Eq.(\ref{Spcomponent}). 
It is obtained in Eqs.(\ref{finalstate+1}-\ref{finalstate-1}) where
 at the initial time $t=t_i$, one-particle state with $l=+1$ is replaced by an anti-neutrino with $l=-1$  and three-particle state with $l=-1$ is replaced by  a neutrino pair plus 
anti-neutrino  with $l=+1$.
For the transitions, it is only necessary to replace $\alpha$ with $\beta$ in Eq.(\ref{Spmatrix}),
\begin{align}
    \begin{pmatrix}
\beta^{\dagger}(\textbf{p},t_f)\ket{0,t_f}_{\bp} \\
B_{\alpha}^{\dagger}(\textbf{p},t_f)\beta^{\dagger}(\textbf{p},t_f)\ket{0,t_f}_{\bp}
\end{pmatrix} &=
\begin{pmatrix}
\mathcal{S}_{\bp}^{11 *} & \mathcal{S}_{\bp}^{13 *} \\
\mathcal{S}_{\bp}^{31 *} & \mathcal{S}_{\bp}^{33 *}
\end{pmatrix}
\begin{pmatrix}
\beta^{\dagger}(\textbf{p},t_i)\ket{0,t_i}_{\bp} \\
B_{\alpha}^{\dagger}(\textbf{p},t_i)\beta^{\dagger}(\textbf{p},t_i)\ket{0,t_i}_{\bp}
\end{pmatrix}.
\end{align}
 Next, we derive the matrix elements for $S_\bq$-matrix operator for the $\bq$ sector, where  transitions from the vacuum  
to states with even lepton numbers take place.   Including the vacuum, 
the eigenvalues of the lepton number for the states 
 in the $\bq$ sector are even numbers. To express the state with lepton number $\pm l$, we use $\ket{\pm l, t }$ while we use $\ket{n, t}$ to denote a $n(>0)$ particle state. 
Note that  the state $\ket{2,t_f}_{\bq}$ defined in Eq.(\ref{21}) is a superposition of $\ket{+2,t_f}_{\bq}$ and $\ket{-2,t_f}_{\bq}$ given as,
\bea
\ket{2,t_f}_{\bq}=\frac{1}{\sqrt{2}}(\ket{+2,t_f}_{\bq}+\ket{-2,t_f}_{\bq}),
\eea
where
$\ket{+2,t}_{\bq}\equiv B_{\alpha}^{\dagger}(\bq,t)\ket{0,t}_{\bq} $ and $\ket{-2,t}_{\bq}\equiv
B_{\beta}^{\dagger}(\bq,t)\ket{0,t}_{\bq}
$.
The time evolution of  $\ket{0,t}_{\bq}$, 
and $\ket{4,t}_{\bq}$ can be derived based on the matrix  $G(\bq,\tau)$ in  Eq.(\ref{TEGmatrix}) and Eq.(\ref{Gm}).
Here  we show the results for $\ket{+2,t_f}_{\bq}$ and $\ket{-2,t_f}_{\bq}$,
\begin{align}\label{eq:+2}
    \ket{+2,t_f}_{\bq} 
    &= e^{2i|\bq|\tau}f(\bq,\tau)g(\bq,\tau)\ket{0,t_i}_{\bq}
    + e^{2i|\bq|\tau}\left|f(\bq,\tau)\right|^2\ket{+2,t_i}_{\bq} \notag \\
    &+ e^{2i|\bq|\tau}\left(-g(\bq,\tau)^2\right)\ket{-2,t_i}_{\bq}
    + e^{2i|\bq|\tau}(-f^*(\bq,\tau)g(\bq,\tau))\ket{4,t_i}_\bq,
\end{align}
\begin{align}\label{eq:-2}
    \ket{-2,t_f}_{\bq} 
    &= e^{2i|\bq|\tau}f(\bq,\tau)g(\bq,\tau)\ket{0,t_i}_{\bq}
    + e^{2i|\bq|\tau}\left(-g(\bq,\tau)^2\right)\ket{+2,t_i}_{\bq} \notag \\
    &+ e^{2i|\bq|\tau}\left|f(\bq,\tau)\right|^2\ket{-2,t_i}_{\bq}
    + e^{2i|\bq|\tau}(-f^*(\bq,\tau)g(\bq,\tau))\ket{4,t_i}_\bq.
\end{align}
For convenience, we  rename the four states  $\ket{0,t}_{\bq}, \ket{+2,t}_{\bq}, \ket{-2,t}_{\bq},\ket{4,t}_{\bq}$ as follows,
\bea
&&\ket{\theta_1,t}_\bq = \ket{0,t}_\bq, \\
&&\ket{\theta_2,t}_\bq = \ket{+2,t}_\bq, \\
&&\ket{\theta_3,t}_\bq = \ket{-2,t}_\bq, \\
&&\ket{\theta_4,t}_\bq = \ket{4,t}_\bq.
\eea
With the states $\ket{\theta_{j_\bq},t}_\bq, (j_\bq=1 \sim 4)$,
 the matrix   $S_\bq^\ast$ relates the states at $t_f$ to those at $t_i$ as follows, 
\begin{align}
   \ket{\theta_{j_\bq},t_f}_\bq = \sum_{k_\bq=1}^4S_{\bq}^{j_\bq k_\bq*}\ket{\theta_{k_\bq},t_i}_\bq \label{Sqmatrix},
\end{align}
where the matrix $S_{\bq}^{*}$ is obtained with Eq.(\ref{TEGmatrix}), Eq.(\ref{Gm}), Eq(\ref{eq:+2}) and Eq.(\ref{eq:-2}),
\begin{align}
S_{\bq}^{*}=
e^{2i|\bq|\tau}\begin{pmatrix}
     f(\bq,\tau)^2 & -f(\bq,\tau)g(\bq,\tau) & -f(\bq,\tau)g(\bq,\tau) & g(\bq,\tau)^2  \\
     f(\bq,\tau)g(\bq,\tau) & |f(\bq,\tau)|^2 & -g(\bq,\tau)^2 & -f^*(\bq,\tau)g(\bq,\tau)  \\
     f(\bq,\tau)g(\bq,\tau) & -g(\bq,\tau)^2 & |f(\bq,\tau)|^2 & -f^*(\bq,\tau)g(\bq,\tau)  \\
     g(\bq,\tau)^2 & f^*(\bq,\tau)g(\bq,\tau) & f^*(\bq,\tau)g(\bq,\tau) & (f^*(\bq,\tau))^2  
    \end{pmatrix}.\label{Sqcomponent}
\end{align}
\subsection{Probability}
\begin{figure}[ht]
\centering
\begin{minipage}[t]{0.49\columnwidth}
    \centering
    \includegraphics[width=0.9\columnwidth]{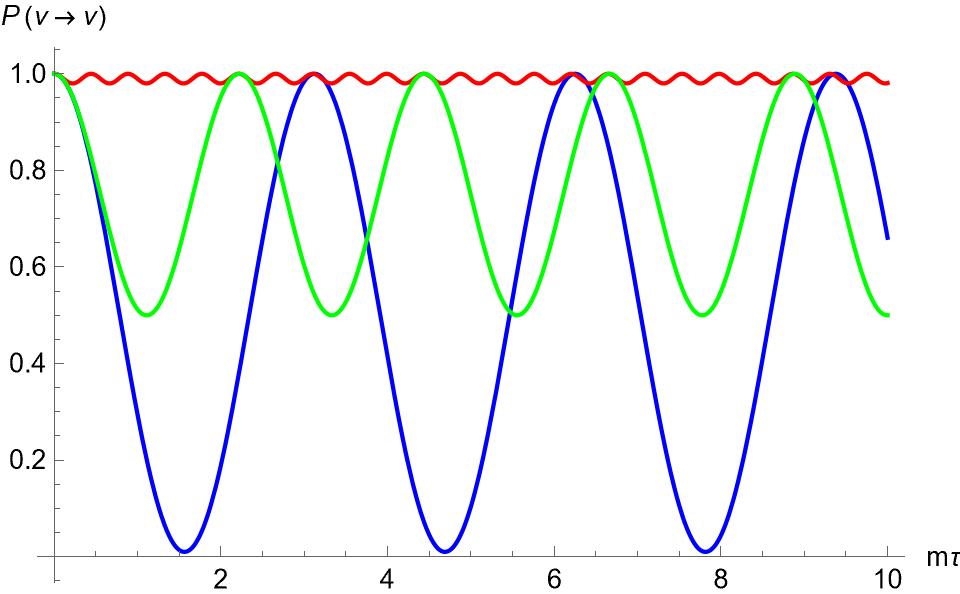}
    \caption{Survival probability: $P_{\nu_\bp \to \nu_\bp}(\bp,\tau)$.}
    \label{fig:P11}
\end{minipage}
\begin{minipage}[t]{0.49\columnwidth}
    \centering
    \includegraphics[width=0.9\columnwidth]{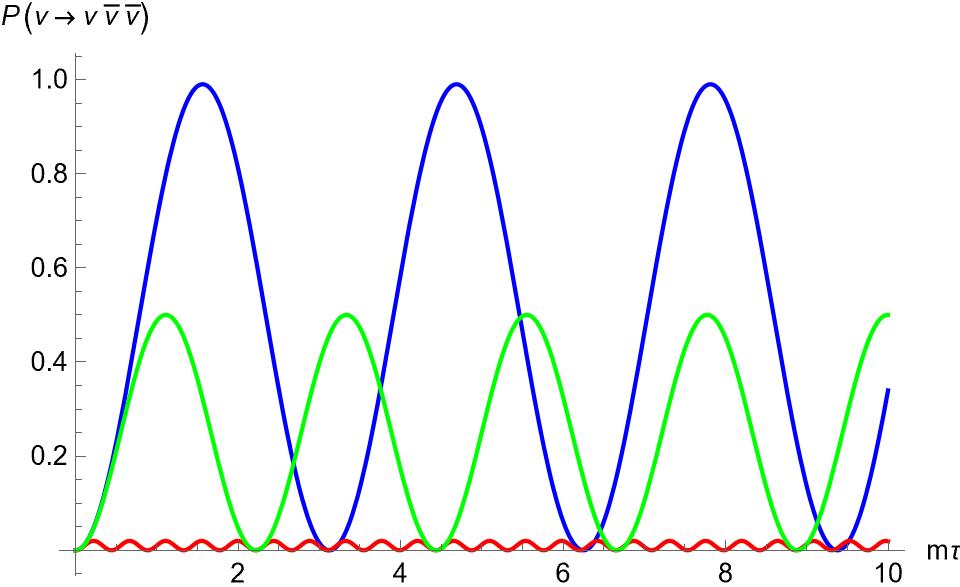}
    \caption{Chiral oscillation probability: $P_{\nu_\bp \to \nu_\bp\bar{\nu}_{-\bp}\bar{\nu}_\bp}(\bp,\tau)$}
    \label{fig:P1M1}
\end{minipage}

\caption*
{ In Fig.\ref{fig:P11}, the survival probability: $P_{\nu_\bp \to \nu_\bp}(\bp,\tau)$ 
and in Fig.\ref{fig:P1M1}, chiral oscillation probability: $P_{\nu_\bp \to \nu_\bp\bar{\nu}_{-\bp}\bar{\nu}_\bp}(\bp,\tau)$
are plotted  as functions of $m \tau$ where $m$ is the mass of the neutrino. The red colored line shows the relativistic case with $v = 0.99$, The green colored line shows the case with $v = \frac{1}{\sqrt{2}}$. The  blue colored line shows the non-relativistic case with $v=0.1$.}
\end{figure}

Based on the matrices $\mathcal{S}_\bp$ and  $S_\bq$, we obtain the survival probability 
 and the chiral oscillation  probability for the neutrino
transitions.
First, we calculate the neutrino transition probability in the $\bp$ sector. From Eq.(\ref{Spmatrix}) and Eq.(\ref{Spcomponent}), the transition probabilities from 
the neutrino with lepton number $l=+1$ to the states with lepton number $l=\pm 1$ in $\bp$ sector are respectively given by,
\begin{align}
\mathcal{P}_{+1 \to +1}(\bp,\tau) &= \left| _{\bp}\langle 0,t_f |\alpha(\bp,t_f)\alpha^{\dagger}(\bp,t_i)|0,t_i\rangle_{\bp} \right|^2, \nn \\
&=  |\mathcal{S}^{11}_\bp|^2, \label{1to1}\\
\mathcal{P}_{+1 \to -1}(\bp,\tau) &= \left| _{\bp}\langle 0,t_f |B_{\beta}(\bp,t_f)\alpha(\bp,t_f)\alpha^{\dagger}(\bp,t_i)|0,t_i\rangle_{\bp} \right|^2, \nn \\
&=  |\mathcal{S}^{31}_\bp|^2.\label{1tom1}
\end{align}
 Next, we calculate the transition probabilities from the vacuum state $|\theta_1,t_i\rangle_{\bq}$ to all the possible states $ |\theta_{j_\bq},t_f \rangle_{\bq}$, ($j_{\bq}=1\sim 4$) in the $\bq$ sector. From Eq.(\ref{Sqmatrix}) and Eq.(\ref{Sqcomponent}), the transition probabilities from the state labelled by $\theta_1$ to $\theta_{j_\bq}$ ($j_\bq=1\sim 4$)  in the $\bq$ sector are,
\begin{align}
&\mathcal{P}_{\theta_1 \to \theta_{j_\bq}}(\bq,\tau) = \left| _{\bq}\langle \theta_{j_\bq},t_f|\theta_1,t_i\rangle_{\bq} \right|^2 = \left|S^{j_{\bq}1}_{\bq} \right|^2. \label{Pq}
\end{align}
The sum of these transition probabilities over the possible final states is given as,
\begin{align}
\sum_{j_{\bq}=1}^4\mathcal{P}_{\theta_1 \to \theta_{j_\bq}}(\bq,\tau)=\sum_{j_{\bq}=1}^4 \left|S^{j_{\bq}1}_{\bq} \right|^2=1,\label{0toall}
\end{align}
which implies the unitarity relation of the vacuum evolution.
 Based on the above, we derive the transition probability taking account of the transitions from all momentum sectors. 
First, from Eq.(\ref{state}), the initial neutrino state with momentum $\bp$  is given as follows, 
\bea
\ket{\nu_\bp \ [j_{\bq}=1], t_i }= \alpha^\dagger(\bp,t_i) \ket{0,t_i}=  \alpha^\dagger(\bp,t_i) \ket{0, t_i}_{\bp} \prod_{\bq \ne \bp} \ket{\theta_1, t_i}_{\bq},  \quad
\bp, \bq \in A,
\label{statep} 
\eea
where $[j_\bq]$ specifies the state of arbitrary $\bq$ sector other than $\bp$ sector and $j_{\bq}$  can take $1 \sim 4$ which corresponds to
the state $\theta_{1} \sim \theta_{4}$. For the initial state in Eq.(\ref{statep}), 
$[j_{\bq}=1]$ implies that the state in all $\bq$ sectors is $\theta_1$.
The possible final states with a neutrino of the momentum $\bp$  to which the transition from the state in Eq.(\ref{statep}) can take place are given as follows,
\bea
\ket{\nu_\bp \ [j_{\bq}], t_f }=\alpha^\dagger(\bp,t_f) \ket{0, t_f}_{\bp} \prod_{\bq \ne \bp} \ket{\theta_{j_{\bq}}, t_f}_{\bq},  \quad j_{\bq}=1\sim 4.
\label{finalp}
\eea
The number of the final states in Eq.(\ref{finalp}) is $4^n$ where 
$n$
 is a possible number of momenta for $\bq$ in A.
Then the transition amplitude from the initial state in Eq.(\ref{statep}) to the final state in Eq.(\ref{finalp}) is given by,
\bea
 \bra{\nu_\bp \ [j_{\bq}] , t_f } \ket{\nu_\bp \ [j_{\bq}=1], t_i } 
&=& \prod_{\bq \ne \bp} \,_\bq \langle \theta_{j_{\bq}},t_f|\, _\bp \langle0, t_f| \alpha(\bp,t_f) \alpha^\dagger(\bp,t_i) |0, t_i\rangle_{\bp} |\theta_1, t_i\rangle_{\bq}, \nn \\
&=&\prod_{\bq \ne \bp}  \,_\bq \langle \theta_{j_{\bq}},t_f|\theta_1, t_i\rangle_{\bq}  \, _\bp \langle0, t_f| \alpha(\bp,t_f) \alpha^\dagger(\bp,t_i) |0, t_i\rangle_{\bp} ,\nn \\
&=&  \prod_{\bq \ne \bp} S^{j_{\bq}1}_{\bq}   \mathcal{S}_{\bp}^{11}.
\label{eachamp}
\eea
When the final state in the $\bp$ sector is a three-particle state, the transition amplitude is given by,
\bea
 \bra{\nu_\bp \bar{\nu}_{-\bp} \bar{\nu}_{\bp} \ [j_{\bq}] , t_f } \ket{\nu_\bp \ [j_{\bq}=1], t_i }
&=& \prod_{\bq \ne \bp} \,_\bq \langle \theta_{j_{\bq}},t_f|\, _\bp \langle0, t_f|B_{\beta}(\bp,t_f) \alpha(\bp,t_f) \alpha^\dagger(\bp,t_i) |0, t_i\rangle_{\bp} |\theta_1, t_i\rangle_{\bq}, \nn \\
&=&\prod_{\bq \ne \bp}  \,_\bq \langle \theta_{j_{\bq}},t_f|\theta_1, t_i\rangle_{\bq}  \, _\bp \langle0, t_f| B_{\beta}(\bp,t_f)\alpha(\bp,t_f) \alpha^\dagger(\bp,t_i) |0, t_i\rangle_{\bp}, \nn \\
&=&  \prod_{\bq \ne \bp} S^{j_{\bq}1}_{\bq}   \mathcal{S}_{\bp}^{31}.
\label{eachampM}
\eea
Then the corresponding transition probability is given by,
\bea
P_{\nu_\bp [j_\bq=1]\to \nu_\bp [j_\bq]}(\bp,\tau) = \prod_{\bq \ne \bp} \mathcal{P}_{\theta_1 \to \theta_{j_\bq}}(\bq,\tau)\mathcal{P}_{+1 \to +1}(\bp,\tau) =\prod_{\bq \ne \bp} |S^{j_{\bq}1}_{\bq}|^2  |\mathcal{S}_{\bp}^{11}|^2,
\label{eachprob}
\eea
\bea
P_{\nu_\bp [j_\bq=1] \to \nu_\bp\bar{\nu}_{-\bp}\bar{\nu}_\bp[j_\bq]}(\bp,\tau) = \prod_{\bq \ne \bp} \mathcal{P}_{\theta_1 \to \theta_{j_\bq}}(\bq,\tau)\mathcal{P}_{+1 \to -1}(\bp,\tau) =\prod_{\bq \ne \bp} |S^{j_{\bq}1}_{\bq}|^2  |\mathcal{S}_{\bp}^{31}|^2.
\label{eachprobM}
\eea
Eq.(\ref{eachprob}) and Eq.(\ref{eachprobM}) are the probabilities when $j_\bq$ in the final state
 is fixed to be one of 1 to 4 at each momentum $\bq$ sector.
Finally we define the survival probability $P_{\nu_\bp \to \nu_\bp}(\bp,\tau)$ as 
 the sum of the probabilities in Eq.(\ref{eachprob}) over the possible final states
specified as $  \ket{ \nu_\bp [j_\bq] \ t_f}$,
\bea
P_{\nu_\bp \to \nu_\bp}(\bp,\tau)&=&
\sum_{[j_\bq]} P_{\nu_\bp [j_\bq=1]\to \nu_\bp [j_\bq]}(\bp,\tau), \nn \\
&=&\prod_{\bq \ne \bp}  \sum_{j_{\bq}=1}^{4} |S^{j_{\bq}1}_{\bq}|^2  |\mathcal{S}_{\bp}^{11}|^2, \nn \\
&=& |\mathcal{S}_{\bp}^{11}|^2=|f(\bp, \tau)|^2, \nn \\
&=&1-(1-v^2)\sin^2E_{\bp}\tau,  \label{P11} 
\label{chip}
\eea
where we use Eq.(\ref{0toall}), Eq.(\ref{Spcomponent}) and Eq.(\ref{eq:f}). $v$ is the velocity defined by $v = \frac{\ap}{E_{\bp}}$.
In the first line of Eq.(\ref{chip}), $\sum_{[j_\bq]}$ denotes the sum over the possible $4^n$ configurations of $j_\bq$ for final states. 
\footnote{The notation $\sum_{[j_\bq]}$ can be understood
as the functional integral over $j_\bq$(a function of $\bq$) which takes one of the integer values $\{1,2,3,4\}$.  Then the probability,.e.g.,
$P_{\nu_\bp [j_\bq=1]\to \nu_\bp [j_\bq]}(\bp,\tau)$ can be understood as the probability of the observing function
$j_{\bq}$ with a neutrino $\nu(\bp)$ as the final state. See  chapter $12$ in \cite{FeynmanHibbs} for the probability of the function $P[f(x)]$ and the functional integral.}
Similarly, we calculate the chiral oscillation probability $P_{\nu_\bp \to \nu_\bp\bar{\nu}_{-\bp}\bar{\nu}_\bp}(\bp,\tau)$ for that the lepton number changing transition occurs from the state with $l_\bp=+1$ to the state with $l_\bp=-1$.
Here $l_\bp$  is the lepton number in $\bp$ sector.
 $P_{\nu_\bp \to \nu_\bp\bar{\nu}_{-\bp}\bar{\nu}_\bp}(\bp,\tau)$ is the sum of Eq.(\ref{eachprobM}) over the possible final states, it is given by,
\begin{align}
P_{\nu_\bp \to \nu_\bp\bar{\nu}_{-\bp}\bar{\nu}_\bp}(\bp,\tau) 
&= \sum_{[j_\bq]} P_{\nu_\bp [j_\bq=1] \to \nu_\bp\bar{\nu}_{-\bp}\bar{\nu}_\bp[j_\bq]}(\bp,\tau) ,\nn \\
&= \prod_{\bq \neq \bp \in A }\sum_{j_{\bq}=1}^4 |S^{j_{\bq}1}_{\bq}|^2  |\mathcal{S}_{\bp}^{31}|^2,  \nn \\
&=  |\mathcal{S}^{31}_\bp|^2 =\left| g(\bp,\tau) \right|^2, \nn \\
&= (1-v^2)\sin^2E_{\bp}\tau, \label{P1M1}
\end{align}
where we use Eq.(\ref{eq:g}), Eq.(\ref{Spcomponent})  and Eq.(\ref{0toall}).
The probabilities of  Eq.(\ref{P11}) and Eq.(\ref{P1M1}) show the oscillation behavior of the two-state system in the $\bp$ sector.  The two states correspond to the lepton number for the $\bp$ sector ; $l_\bp=\pm 1$ in  Eq.(\ref{1to1}) and Eq.(\ref{1tom1}),
\bea
\ket{l_\bp=+1,t}&=& \ket{\nu_\bp, t}\equiv \alpha^\dagger(\bp,t) |0, t \rangle_{\bp} , \\
\ket{l_\bp=-1,t}&=&\ket{\nu_\bp\bar{\nu}_{-\bp}\bar{\nu}_\bp, t}\equiv \alpha^\dagger(\bp,t) B^\dagger_\beta(\bp,t)|0, t \rangle_{\bp}.
\eea
We also note that 
the survival probability $P_{\nu_{\bp} \to \nu_{\bp}}(\bp,\tau)$ and the chiral oscillation probability $P_{\nu_\bp \to \nu_\bp\bar{\nu}_{-\bp}\bar{\nu}_\bp}(\bp,\tau)$, are equal to the lepton number conserving probability $\mathcal{P}_{+1 \to +1}(\bp,\tau)$ in Eq.(\ref{1to1}) and the lepton number violating probability $\mathcal{P}_{+1 \to -1}(\bp,\tau)$ in Eq.(\ref{1tom1}), respectively as,
\begin{align}
P_{\nu_{\bp} \to \nu_{\bp}}(\bp,\tau)=\mathcal{P}_{+1 \to +1}(\bp,\tau) ,\\
P_{\nu_{\bp} \to \nu_\bp \bar{\nu}_{-\bp}\bar{\nu}_\bp}(\bp,\tau)=\mathcal{P}_{+1 \to -1}(\bp,\tau).
\end{align}
These probabilities satisfies the unitarity relation,
\bea
P_{\nu_{\bp} \to \nu_{\bp}}(\bp,\tau)+P_{\nu_{\bp} \to \nu_\bp \bar{\nu}_{-\bp}\bar{\nu}_\bp}(\bp,\tau)=1.
\label{unitary}
\eea
This unitarity relation is similar to Eq.(54) in \cite{Blasone:2024zsn}. However, the definition of each probability is different from theirs.  In Appendix \ref{sec:Blasone}, we explain the difference of probabilities between our approach and the perturbative approach in Ref.\cite{Blasone:2024zsn}.
Below we explain how the oscillation behaviour of the probabilities occurs. For simplicity, we assume the neutrino velocity is very small $0 < v \ll 1$.
See the blue curves of Fig.\ref{fig:P11} and Fig.\ref{fig:P1M1} where the probabilities for $v=0.1$ are shown.
At $\tau=0$ i.e., $t=t_i$, 
\bea
(P_{\nu_{\bp} \to \nu_{\bp}}(\bp,0), P_{\nu_{\bp} \to \nu_\bp \bar{\nu}_{-\bp} \bar{\nu}_\bp}(\bp,0)) =(1,0).
\eea
In the small velocity limit, the probabilities in Eqs.(\ref{chip}-\ref{P1M1}) oscillate as,
\bea
(P_{\nu_{\bp} \to \nu_{\bp}}(\bp, \tau), P_{\nu_{\bp} \to \nu_\bp \bar{\nu}_{-\bp} \bar{\nu}_\bp}(\bp, \tau))=(1-\sin^2 m \tau,  \sin^2 m\tau).
\label{exact}
\eea 
As the time $\tau$ increases, the Majorana mass term creates the anti-neutrino pair $\bar{\nu}(\bp) \bar{\nu}(-\bp)$. Then it produces to the three particle final state  
in $\bp$ sector of the upper right  in Fig.\ref{each sectors}.  For the neutrino with the small non-zero velocity
$0< v \ll 1$, it is sufficient to consider the Majorana mass terms in the Hamiltonian $h(\bp)$ in Eq.(\ref{16}) as $|\bp| \ll m$,
\begin{align}
    h_m(\bp,t_i) 
    &=-im[B_{\alpha}(\bp,t_i)+B_{\beta}(\bp,t_i) - B^{\dagger}_{\alpha}(\bp,t_i)-B^{\dagger}_{\beta}(\bp,t_i)] \label{hm}.
\end{align}
The amplitude for the creating $\bar{\nu}(\bp) \bar{\nu}(-\bp)$ from the vacuum in the infinitesimal time interval $\delta \tau$ is proportinal to $m \delta \tau$.
Then the probabilities for finding the $l_\bp=\pm 1$ states are given up to the second order of $m$ by,
\bea
(P_{\nu_{\bp} \to \nu_{\bp}}(\bp,\delta \tau), P_{\nu_{\bp} \to \nu_\bp \bar{\nu}_{-\bp} \bar{\nu}_\bp}(\bp, \delta \tau))=(1-m^2 (\delta \tau)^2, m^2 (\delta \tau)^2).
\eea 
which is consistent with the exact formula in Eq.(\ref{exact}).
At the beginning $m \delta \tau <1$, the transition probability from $l_{\bp}=+1$ to $l_\bp=-1$  increases and the survival probability decreases.  Then  eventually $P(l_\bp=1 \to l_\bp=1, \tau)$ becomes smaller than
$P(l_\bp=1 \to l_\bp=-1, \tau)$. When the $l_\bp=-1$ state that includes $\bar{\nu}(\bp) \bar{\nu}(-\bp)$ is dominant and the probability for the state with  $l_\bp=1$   becomes negligible,  an anti-Majorana neutrino pair $\bar{\nu}(\bp) \bar{\nu}(-\bp)$
in $l_\bp=-1$ state begins to be annihilated into the vacuum by Majorana mass term.
 Then
 the probability for the state with $l_\bp=1$ starts to increase. 
In this way, the probabilities for $l_\bp=1$ and $l_\bp=-1$ oscillates
decreasing and increasing alternately by keeping the unitary constraint Eq.(\ref{unitary}).

 In Fig.\ref{fig:P11}, the survival probability, i.e., Eq.(\ref{P11}) and in Fig.\ref{fig:P1M1}, the chiral oscillation probability, i.e., Eq.(\ref{P1M1}) are respectively plotted for various velocities of the neutrino. 
The former oscillates between $[v^2, 1]$ and the latter oscillates between $[0, 1-v^2]$.
In the case of relativistic neutrino with $v \approx 1$, The survival probability stays around $1$ and the chiral oscillation probability remains close to $0$ with a short period $ \Delta \tau \simeq \frac{0.1 \sqrt{2} \pi}{m}$ for $v=0.99$.  In the case of non-relativistic neutrino with $0 <v \ll1$, both the survival and chiral oscillation probabilities oscillate between $0$ and $1$  with a long period 
$\Delta \tau \simeq \frac{\pi}{m}$.
\section{Conclusion}\label{sec:Conclusion}
We study the probability for chiral oscillation of Majorana neutrino in quantum field theory. In this paper, we first show that the Hamiltonian can be written without the zero momentum mode. Secondly,  we show that the state developed from the vacuum state is a superposition of the vacuum state, the two-particle state, and the four-particle state and the time evolution can be described as the Bogoliubov transformation.   Furthermore, the state with more than four particles in an arbitrary momentum sector are not allowed 
due to Pauli exclusion principle. 
 For the neutrino oscillation probability in this work, if neutrino is ultra-relativistic, i.e., $v \approx 1$, the chirality flip is suppressed.  In other words, the chirality flip occurs for the non-relativistic case.  Moreover, we find that  the chiral oscillation is not neutrino and anti-neutrino oscillation, but a transition from a neutrino state to the state with an anti-neutrino Cooper pair plus a neutrino (three-particle state). This is because the Majorana mass term creates anti-neutrino Cooper pair from the vacuum and it appears through time. 
The chiral transition to the final three-particle state
with $l=-1$ is different from the transition to a single anti-neutrino state.
One can show that neutrino and anti-neutrino oscillation do not occur using operators and eigenstates with definite lepton numbers.
In this work, we show the former transition indeed occurs during the time evolution of a neutrino state.
In Appendix \ref{sec:appendix proof}, we show that the probability for neutrino and anti-neutrino oscillation vanishes in the mass basis as well.

Lastly, we compare our results with the previous study
\cite{Morozumi:2022mqh} on the expectation value of  lepton number operator.
The expectation value of the lepton number operator in the work is, 
\bea
\bra{\nu(\bp,t_i)}L(\bp,t_f)\ket{\nu(\bp,t_i)} 
&= v^2 +(1-v^2)\cos(2E_{\bp}\tau).  
\label{eq:lexp}
\eea
In the present work, the method to compute the probabilities is developed. Using the probabilities,
one can obtain the expectation value of the lepton number in $\bp$ sector in another way, i.e.,
the expectation value equals to the difference of the survival probability for 
 $l=+1$ to $l=+1$ and the chiral oscillation probability for the transition from $l=+1$ to $l=-1$,
\begin{align}
 P_{\nu_\bp \to \nu_\bp}(\bp,\tau)-P_{\nu_\bp \to \nu_\bp\bar{\nu}_{-\bp}\bar{\nu}_\bp}(\bp,\tau) 
&=1-2(1-v^2) \sin^2 E_{\bp} \tau,
\label{Leptonnumber}
\end{align}  
 where we use the result
of Eq.(\ref{P11}) and Eq.(\ref{P1M1}).   We emphasize this is the expectation value of the lepton number of the $\bp$ sector at  $t_f$ since
$  P_{\nu_\bp \to \nu_\bp}(\bp,\tau)$ is the probability of finding  $\nu_\bp$ with $l=+1$
and  $P_{\nu_\bp \to \nu_\bp \bar{\nu}_{-\bp}\bar{\nu}_\bp}(\bp,\tau) $ is the probability of  finding  $ \nu_\bp \bar{\nu}_{-\bp}\bar{\nu}_\bp$ with $l=-1$
at  $\tau=t_f-t_i$.
Indeed, the results of Eq.(\ref{eq:lexp}) and Eq.(\ref{Leptonnumber}) are the same to each other. 
This shows that the probabilities obtained in the present framework lead to a consistent result for the expectation value of the lepton number operator; Fig.1 of
\cite{Morozumi:2022mqh}.   For the relativistic  neutrino  ($v \lessapprox1$), it behaves as $\simeq 1-4 (1-v)\sin^2 \frac{m}{\sqrt{1-v^2}} \tau \simeq 1$ 
while for non-relativistic neutrino ($v \ll 1$), it oscillates as $v^2+(1-v^2)
\cos 2E_p \tau
\simeq \cos \frac{2 m \tau }{\sqrt{1-v^2}} $  between $+1$ and $-1$.

In future work, we will expand this formula to the three-flavor case. Furthermore, we need to discuss how the oscillation probability is affected when the matter effects are considered
\cite{Li:2023iys, Cirigliano:2024pnm}.
\section*{Acknowledgement}
We would like to thank Naoki Uemura for the useful comment and discussion.  
\begin{appendices} 
\renewcommand{\thesection}{\Alph{section}}
\counterwithin*{equation}{section}
\renewcommand\theequation{\thesection\arabic{equation}}
\section{Derivations of the Hamiltonian and anti-commutation relations without  zero mode}
\label{sec:appendix zeromode}
In this appendix, we derive the Hamiltonian, Eq.(\ref{Hamiltonianwozero}) for the Majorana neutrino without zero mode
 operator.  We also derive the anti-commutation
relations of the field operators $\eta, \eta^\dagger$ Eqs.(\ref{acrelations1}-\ref{acrelations2})
based on the calculation of the Dirac bracket \cite{Dirac1964}.
As a result, the anti-commutation relation becomes a modified Dirac delta function without zero 
mode as in Eq.(\ref{acrelations1}). 
This enables us to expand the Majorana field operator with the plane wave massless spinors.
 Then the  Hamiltonian is also expanded by 
with creation and annihilation operators with non-zero momentum as given in Eq.(\ref{eq:Hamiltonian}). 

From the Lagrangian in Eq.(\ref{Lagrangianp}), the conjugate momentum for each field is,
\begin{align}
\pi_\eta = \frac{\partial \mathcal{L}'}{\partial \dot{\eta}} =i\eta^\dagger,\quad \pi_{\eta^\dagger} =0,\quad \pi_{\xi_0}=\pi_{\xi_0^\dagger}=0.
\end{align}
  Then we obtain four constraints:
\begin{align}
\phi^1(x)=\pi_\eta-i\eta^\dagger,\quad \phi^2(x)=\pi_{\eta^\dagger},\quad \phi^3=\pi_{\xi_0},\quad \phi^4=\pi_{\xi_0^\dagger}.
\end{align} 
The Hamiltonian density is given by,
\begin{align}
\mathcal{H}'' &= (\pi_\eta-i\eta^\dagger)\dot{\eta} + \pi_{\eta^\dagger}\dot{\eta^\dagger}+\pi_{\xi_0}\dot{\xi_0}+\pi_{\xi_0^\dagger}\dot{\xi_0^\dagger} \nn \\
&+\eta^{\dagger}i\boldsymbol{\sigma} \cdot \boldsymbol{\nabla}\eta +\frac{m}{2}(- \eta^{\dagger}i\sigma_2\eta^\dagger + \eta i\sigma_2\eta)+\frac{i}{V}(\xi_0^\dagger \eta_0-\eta_0^\dagger \xi_0).
\end{align}
We add the constraint to the Hamiltonian with the help of Lagrange multiplier,
\begin{align}
\mathcal{H}' = \mathcal{H} +\sum_{A=1}^4 \phi^{A}\lambda^{A}.
\end{align}
$\mathcal{H}$ is
\begin{align}
\mathcal{H}=\eta^{\dagger}i\boldsymbol{\sigma} \cdot \boldsymbol{\nabla}\eta +\frac{m}{2}(- \eta^{\dagger}i\sigma_2\eta^\dagger + \eta i\sigma_2\eta)+\frac{i}{V}(\xi_0^\dagger \eta_0-\eta_0^\dagger \xi_0),
\label{H0density}
\end{align}
where we absorb the velocity dependent terms into the constraints by redefining the Lagrange multipliers,
\begin{align}
\lambda^1 \to \lambda^1-\dot{\eta},\quad \lambda^2 \to \lambda^2-\dot{\eta^\dagger},\quad \lambda^3 \to \lambda^3 -\dot{\xi_0},\quad \lambda^4 \to \lambda^4 -\dot{\xi_0^\dagger}.
\end{align}
Then  one can consider the Hamiltonian given below,
\bea
H' &\equiv&\int d^3\bx \left[ \mathcal{H} +\sum_{a=1}^2 \phi^{a}(x)\lambda^{a}(x) \right]+\sum_{\alpha=3}^4 \phi^{\alpha}\lambda^{\alpha}, \\
&=& H + \sum_{a=1}^2 \int d^3\bx \phi^{a}(x)\lambda^{a}(x) 
+\sum_{\alpha=3}^4 \phi^{\alpha}\lambda^{\alpha}, \\
H &\equiv&  \int d^3\bx  \mathcal{H}.
\eea
 where we replace $\lambda^\alpha $ with $\frac{\lambda^\alpha}{V}$ for $\alpha=3,4$ to impose the position independent constraints $\phi^\alpha=0 (\alpha=3,4)$ on the Hamiltonian from those on the Hamiltonian density.
Next we examine if  the constraints $\phi^a(x)=0 \ (a=1,2)$ and 
$\phi^{\alpha}=0 \ (\alpha=3,4)$ do not contradict with the time evolution,
\begin{align}
\dot{\phi}^a(x) = \{\phi^a(x),H\}_{PB} +\int d^3\by\sum_{b=1}^2 \{\phi^a(x),\phi^b(y)\}_{PB}\lambda^b(y)+\sum_{\beta=3}^4 \{ \phi^a(x),\phi^{\beta} \}_{PB}\lambda^{\beta} \approx 0,
\end{align}
where PB denotes the Poisson bracket. One obtains the conditions,
\begin{align}
\begin{pmatrix}\dot{\phi}^1(x) \\\dot{\phi}^2(x) \end{pmatrix} 
=\begin{pmatrix}\{\phi^1(x),H\}_{PB} \\ \{\phi^2(x),H\}_{PB} \end{pmatrix} -i\begin{pmatrix}\lambda^2(x) \\\lambda^1(x) \end{pmatrix} \simeq 0.
\end{align}
Then the Lagrange multipliers $\lambda^{1,2}(x)$ can be determined so that the constraints $\phi^{a}(x)=0$ ($a=1,2$) are consistent with the time evolution. 
We also require the  other constrants $ {\phi}^{\alpha} (\alpha=3,4)$ are consistent with time evolution,
\begin{align}
\dot{\phi}^{\alpha} &= \{\phi^{\alpha},H\}_{PB} +\int d^3\by\sum_{b=1}^2 \{\phi^{\alpha},\phi^b(y)\}_{PB}\lambda^b(y)+\sum_{\beta=3}^4 \{ \phi^{\alpha},\phi^{\beta} \}_{PB}\lambda^{\beta}, \nn \\
&= \{\phi^{\alpha},H\}_{PB} \approx 0.
\end{align}
Then we obtain the secondary constraints,
\begin{align}
 \{\phi^{3},H\}_{PB} = i\eta_0^\dagger  \approx 0, \label{constraint5} \\
 \{\phi^{4},H\}_{PB} = i\eta_0 \approx 0. \label{constraint6}
\end{align}
We add the secondary constraints in Eqs.(\ref{constraint5}-\ref{constraint6}) 
to the Hamiltonian,
\begin{align}
H' = \int d^3\bx \left[ \mathcal{H} +\sum_{a=1}^2 \phi^{a}(x)\lambda^{a}(x) \right]+\sum_{\alpha=3}^4 \phi^{\alpha}\lambda^{\alpha} + \sum_{\alpha=5}^6 \phi^{\alpha}\lambda^{\alpha}.
\label{renameH}
\end{align}
In Eq.(\ref{renameH}),
we rename  the constrains as $\phi^3\equiv \eta_0$, $\phi^4 \equiv \eta^\dagger_0$, $\phi^5 \equiv \pi_{\xi_0}$ and $\phi^6 \equiv \pi_{\xi_0^\dagger}$ respectively
for the later convenience.
With the introduction of the new constraints, $\dot{\phi^1}(x)$ is  given  by, as follows,
\begin{align}
\dot{\phi}^1(x) &= \{\phi^1(x),H\}_{PB} +\int d^3\by\{\phi^1(x),\phi^2(y)\}_{PB}\lambda^2(y)+\sum_{\beta=3}^6 \{ \phi^1(x),\phi^{\beta} \}_{PB}\lambda^{\beta}, \nn \\
&=  \{\phi^1(x),H\}_{PB} -i\lambda^2(x)+\frac{1}{V}\lambda^3.
\label{phi1dot}
\end{align}
Similarly, $\dot{\phi^2}(x)$ is given as follows,
\begin{align}
\dot{\phi}^2(x) 
&=  \{\phi^2(x),H\}_{PB} -i\lambda^1(x)+\frac{1}{V}\lambda^4.
\label{phi2dot}
\end{align}
In addition, the time derivative of the new constraints $\phi^3=\eta_0$ and $\phi^4=\eta^\dagger_0$ are
\begin{align}
\dot{\phi}^{3} &= \{\phi^{3},H\}_{PB} +\int d^3\by\sum_{b=1}^2 \{\phi^{3},\phi^b(y)\}_{PB}\lambda^b(y)+\sum_{\beta=3}^6 \{ \phi^{3},\phi^{\beta} \}_{PB}\lambda^{\beta},\nn \\
&=  \{\phi^{3},H\}_{PB} + \frac{1}{V}\int d^3\bx\lambda^1(x),
\label{phi3dot}
\end{align}
and
\begin{align}
\dot{\phi}^{4} &= \{\phi^{4},H \}_{PB} + \frac{1}{V}\int d^3\bx\lambda^2(x).
\label{phi4dot}
\end{align}
$\dot{\phi}^5$ and $\dot{\phi}^6$ do not change from Eqs.(\ref{constraint5}-\ref{constraint6}),
\begin{align}
&\dot{\phi}^{5}=\{\phi^{5}, H'\}_{PB}=\{\phi^{5},H\}_{PB} = i \phi^4= i \eta^\dagger_0, \label{constraint5p} \\
&\dot{\phi}^{6}=\{\phi^{6}, H'\}_{PB}=\{\phi^{6},H\}_{PB} = i \phi^3= i \eta_0. \label{constraint6p}
\end{align}
However the Lagrange multipliers $\lambda^5$ and $\lambda^6$ are undetermined from the conditions $\dot{\phi}^{A}=0, A=1\sim6$ which we have examined.
For these primary constraints, we impose the gauge-fixing like conditions $\phi^7=\phi^8=0$ as constraints,
\begin{align}
\phi^7 = \xi_0\quad, \quad \phi^8=\xi_0^\dagger.
\end{align}
Including all the constraints in Table \ref{constraints}, Hamiltonian is,
\begin{align}
H' = \int d^3\bx \left[ \mathcal{H} +\sum_{a=1}^2 \phi^{a}(x)\lambda^{a}(x) \right]+\sum_{\alpha=3}^4 \phi^{\alpha}\lambda^{\alpha} + \sum_{\alpha=5}^6 \phi^{\alpha}\lambda^{\alpha} + \sum_{\alpha=7}^8 \phi^{\alpha}\lambda^{\alpha}.
\end{align}

Below we show the obtained constraints become the second class by determining the Lagrange multipliers. 
We first consider $\dot{\phi^\alpha}=0 \ (\alpha=5 \sim 8)$,
\begin{align}
\dot{\phi}^{5} &= \{\phi^{5},H\}_{PB} + \sum_{\beta=5}^8 \{ \phi^{5},\phi^{\beta} \}_{PB}\lambda^{\beta}, \nn \\
&=  \{\phi^{5},H\}_{PB} +\{ \phi^{5},\phi^{7} \}_{PB}\lambda^{7}, \nn \\
&=i \phi^4+\lambda^{7} \simeq 0, \\
& \lambda^{7}=-i \eta^\dagger_0,
\end{align}
similarly, one obtains,
\begin{align}
\dot{\phi}^{6} &=\{\phi^{6},H\}_{PB} +\lambda^{8}, \nn \\
&= i \phi^{3} +\lambda^{8}\simeq 0, \\
\lambda^{8}&=-i \eta_0,
\end{align}
\begin{align}
\dot{\phi}^{7} &=\{\phi^{7},H\}_{PB} +\lambda^{5},\nn \\
&=\lambda^{5}\simeq 0, \\
& \lambda^{5}=0,
\end{align}
\begin{align}
\dot{\phi}^{8} &=\{\phi^{8},H\}_{PB} +\lambda^{6}, \nn \\
&=\lambda^{6}\simeq 0, \\
& \lambda^{6}=0.
\end{align}
In this way, $(\lambda^5, \lambda^6,\lambda^7, \lambda^8)$ are determined
as,
\bea
\begin{pmatrix} \lambda^5 & \lambda^6 & \lambda^7 & \lambda^8 \end{pmatrix}
= \begin{pmatrix} 0 & 0 & -i \eta^\dagger_0 & -i \eta_0 \end{pmatrix}.
\label{eq:l5678}
\eea
The other multipliers $\lambda^1(x), \lambda^2(x), \lambda^3$ and $\lambda^4$
are determined as follows. 
We first define zero modes $(\lambda_0^1,\lambda_0^2) $ of the Lagrange multipliers $(\lambda^1(x), \lambda^2(x))$ as,
\begin{align}
\lambda_0^1 &= \frac{1}{V}\int d^3\bx\lambda^1(x), \quad
\lambda_0^2 = \frac{1}{V}\int d^3\bx\lambda^2(x).
\end{align}
With $\{ \phi^a, H \}_{PB} \simeq 0$ $(a=3,4)$ in Eqs.(\ref{phi3dot}-\ref{phi4dot}), the conditions $ \dot{\phi^a}=0 $ $(a=3,4)$ lead to,
\bea
\lambda_0^1=\lambda_0^2=0.\label{eq:l10l20}
\eea
We also define the zero modes $\phi^a (a=1,2)$ of $\phi^1(x)$ and $\phi^2(x)$,
\begin{align}
\phi^1 &= \int d^3\bx\phi^1(x), \quad
\phi^2 = \int d^3\bx \phi^2(x).
\end{align}
One subtracts the zero modes from $\phi^a(x) (a=1,2)$ and defines $\phi'^a(x) (a=1,2)$,
\begin{align}
\phi'^1(x) &=\phi^1(x) - \frac{\phi^1}{V}, \quad
\phi'^2(x) = \phi^2(x) - \frac{\phi^2}{V}.
\end{align}
which satisfy $\int d^3 \bx \phi'^a(x)=0$. Lastly, the conditions $\dot{\phi}^{a}(x)=0, \ (a=1,2)$ in 
Eqs.(\ref{phi1dot}-\ref{phi2dot}) and
\bea
&& \{\phi^1(x), H \}_{PB}=i \left(  \boldsymbol{\nabla} \eta^\dagger(x) \cdot \boldsymbol{\sigma}  + m\sigma_2 \eta(x)-\frac{1}{V}\xi_0^\dagger
\right), \\
&& \{\phi^2(x), H \}_{PB}=i \left( \boldsymbol{\sigma} \cdot \boldsymbol{\nabla} \eta(x) - m\sigma_2 \eta^\dagger(x)-\frac{1}{V}\xi_0 \right), 
\eea
lead to the conditions,
\begin{align}
i \left(  \boldsymbol{\nabla} \eta^\dagger(x) \cdot \boldsymbol{\sigma}  + m\sigma_2 \eta(x)-\frac{1}{V}\xi_0^\dagger
\right)
-i\lambda^2(x)+\frac{1}{V}\lambda^3=0,
\label{phi1dot0}
\end{align}
\begin{align}
&
i \left( \boldsymbol{\sigma} \cdot \boldsymbol{\nabla} \eta(x) - m\sigma_2 \eta^\dagger (x)-\frac{1}{V}\xi_0 \right)
 -i\lambda^1(x)+\frac{1}{V}\lambda^4=0.
\label{phi2dot0}
\end{align}
By integrating Eq.(\ref{phi1dot0}) and Eq.(\ref{phi2dot0}) over three dimentional space $\bx$,
one determines $\lambda^3$ and $\lambda^4$ as,
\bea
\lambda^3&=& i V \lambda^2_0 -i (m \sigma_2 V \eta_0-\xi^\dagger_0)=-i (m \sigma_2 V \eta_0-\xi^\dagger_0), \label{eq:l3}\\
\lambda^4&=& i V \lambda^1_0 -i (-m \sigma_2 V \eta^\dagger_0-\xi_0)=-i (-m \sigma_2 V \eta^\dagger_0-\xi_0), \label{eq:l4}
\eea
where we use Eq.(\ref{eq:l10l20}).
By substituting Eq.(\ref{eq:l3}) to Eq.(\ref{phi1dot0}) and Eq.(\ref{eq:l4}) to Eq.(\ref{phi2dot0}) respectively,
one determines $\lambda^1(x)-\lambda_0^1$ and $\lambda^2(x)-\lambda_0^2$. Including them,
the Lagrange multipliers determined from the conditions of $\dot{\phi}^{A}=0, A=1\sim8$ are summarized as follows,
\begin{align}
&\lambda^1(x)-\lambda_0^1 = \boldsymbol{\sigma} \cdot \boldsymbol{\nabla} \eta(x) -  m\sigma_2(\eta^\dagger(x)-\eta_0^\dagger), \\
&\lambda^2(x)-\lambda_0^2=   \boldsymbol{\nabla} \eta^\dagger(x) \cdot \boldsymbol{\sigma} + m\sigma_2(\eta(x)-\eta_0), \\
&\lambda_0^1 = 0, \\
&\lambda_0^2 = 0,\\
&\lambda^3 = -i\sigma_2mV\eta_0 + i \xi^\dagger_0,\\
&\lambda^4 = i\sigma_2mV\eta_0^\dagger +i \xi_0,\\
&\lambda^5 = 0, \\
&\lambda^6 = 0, \\
&\lambda^7 =  -i\eta_0^\dagger, \\
&\lambda^8 =  -i\eta_0,
\end{align}
where we show  Eq.(\ref{eq:l5678}), Eq.(\ref{eq:l10l20}), Eq.(\ref{eq:l3}), and Eq.(\ref{eq:l4}) again. This shows that the constraints are second class.

We obtain anti-commutation relations among $\eta$ and $\eta^\dagger$  based on the Dirac brackets.
 The Dirac bracket for $\eta$ and $\eta^\dagger$ is calculated as,
\begin{align}
\{\eta(\bx,t),\eta^{\dagger}(\by,t)\}_{DB} &= \{\eta(\bx,t),\eta^{\dagger}(\by,t)\}_{PB} \nn \\
&-\sum_{a,b=1}^{2}\int d^3\bx'd^3\by'\{\eta(\bx,t),\phi'^a(\bx',t)\}_{PB}(\bar{C}^{-1})^{ab}(\bx',\by')\{\phi'^b(\by',t),\eta^{\dagger}(\by,t)\}_{PB} \nn \\
&-\sum_{\alpha,\beta=1}^{4}\{\eta(\bx,t),\phi^{\alpha}\}_{PB}(C_0^{-1})^{\alpha \beta}\{\phi^{\beta},\eta^{\dagger}(\by,t)\}_{PB} \nn \\
&-\sum_{\alpha,\beta=5}^{8}\{\eta(\bx,t),\phi^{\alpha}\}_{PB}(C_{\xi}^{-1})^{\alpha \beta}\{\phi^{\beta},\eta^{\dagger}(\by,t)\}_{PB}
\end{align}
where DB denotes the Dirac bracket.  We define the matrices of the Poisson brackets for
constraints and their inverse matrices,
\begin{align}
\bar{C}^{ab}(\bx, \by)&=\{\phi'^a(\bx), \phi'^b(\by) \}_{PB} = \begin{pmatrix}0 & -i \\
-i & 0\end{pmatrix} \times \left(\delta^{(3)}(\bx-\by) -\frac{1}{V} \right), \ a,b=1\sim 2,\\
C_0^{\alpha\beta}&= \{\phi^\alpha, \phi^\beta\}_{PB} = \begin{pmatrix}0 & -i V & 1 & 0\\
-iV & 0 & 0 & 1 \\
1 & 0 & 0 & 0 \\
0 & 1 & 0 & 0 \end{pmatrix}, \alpha,\beta=1\sim 4,\\
C_{\xi}^{\alpha\beta} &= \{\phi^\alpha, \phi^\beta\}_{PB}=\begin{pmatrix}0 & 0 & 1 & 0\\
0 & 0 & 0 & 1 \\
1 & 0 & 0 & 0 \\
0 & 1 & 0 & 0 \end{pmatrix}, \ \alpha,\beta=5\sim 8.
\end{align}
Each inverse matrix is calculated to be,
\begin{align}
(\bar{C}^{-1})^{ab}(\bx,\by) &= \begin{pmatrix}0 & i \\
i & 0\end{pmatrix} \times \left(\delta^{(3)}(\bx-\by) -\frac{1}{V} \right), \ a,b=1\sim2,\\
(C_0^{-1})^{\alpha \beta} &= \begin{pmatrix}
0 & 0 & 1 & 0\\
0 & 0 & 0 & 1 \\
1 & 0 & 0 &  iV \\
0 & 1 & iV & 0 \end{pmatrix}, \ \alpha,\beta=1\sim 4, \\
(C_{\xi}^{-1})^{\alpha\beta} &= \begin{pmatrix}0 & 0 & 1 & 0\\
0 & 0 & 0 & 1 \\
1 & 0 & 0 & 0 \\
0 & 1 & 0 & 0 \end{pmatrix}, \ \alpha,\beta=5 \sim 8.
\end{align}
Then the Dirac bracket between $\eta$ and $\eta^\dagger$ are given as,
\begin{align}
\{\eta(\bx,t),\eta^{\dagger}(\by,t)\}_{DB} &= -i(\delta^{(3)}(\bx-\by)-\frac{1}{V}) \equiv -i\bar{\delta}^{(3)}(\bx-\by).
\end{align}
One can also show the following Dirac brackets,
\begin{align}
\{\eta(\bx,t),\eta(\by,t)\}_{DB} &= \{\eta^{\dagger}(\bx,t),\eta^{\dagger}(\by,t)\}_{DB} = 0.
\end{align}
Then the anti-commutation relations are,
\begin{align}
\{\eta(\bx,t),\eta^{\dagger}(\by,t)\} &= \delta^{(3)}(\bx-\by)-\frac{1}{V} \label{acrelations3},\\
\{\eta(\bx,t),\eta(\by,t)\} &= \{\eta^{\dagger}(\bx,t),\eta^{\dagger}(\by,t)\} = 0.  \label{acrelations4} 
\end{align}
Thus, the field operators $\eta$ and $\eta^\dagger$ satisfy an anti-commutation relation excluding zero mode because $\{\eta_0, \eta_0^\dagger\}
=\{\eta_0, \eta_0\}=\{\eta_0^\dagger, \eta_0^\dagger \}=0$.

\section{Difference of probabilities between the present approach and the perturbative approach in Ref. \cite{Blasone:2024zsn}} \label{sec:Blasone}
 We note the similarity between our unitarity relation in Eq.(\ref{unitary}) and the unitarity relation of Blasone et al. in Eq.(54) of \cite{Blasone:2024zsn}. 
In this appendix, we explain the difference between their definition of probabilities and ours.
In their paper, they studied the flavor oscillation probability
with perturbative approximation on the off-diagonal elements of Dirac mass matrix.
On the other hand, we focus on one-flavor case of the Majorana neutrino and obtain the oscillation probability in a non-perturbative approach.
In spite of the difference, the unitarity relations in both approaches are similar.
The unitarity relation for the one-flavor case can be expressed as follows, 
\bea
P(\nu(\bp) \to \nu(\bp))+\sum_{X \ne \nu(\bp)}P(\nu(\bp)\rightarrow X)=1,
\label{Blasonelike}
\eea
where the first term corresponds to what they call as the survival probability $P_S$ and the second term as the decay probability $P_D$.
Then, the process $\nu(\bp)\rightarrow X$ includes all the processes except the process $\nu_\bp \to \nu_\bp$ in the $\bp$ sector and $\theta_{1} \rightarrow \theta_{1}$ in 
all the $\bq$ sectors.
Below we compare their survival probability with ours. We also compare their decay probability with our chiral oscillation probability.
One can write their survival probability and  the decay probability
 using our notation,
\bea
P_S&=&P(\nu(\bp) \to \nu(\bp)),\nn \\
    &=&P_{\nu_{\bp} \rightarrow \nu_{\bp}}(\bp,\tau) \prod_{\bq \ne \bp} P_{\theta_{1} \rightarrow \theta_{1}}(\bq,\tau),
\label{BPS}
 \\
P_D&=&P_{\nu_{\bp} \rightarrow \nu_{\bp} \bar{\nu}_{\bp} \bar{\nu}_{-\bp}}(\bp,\tau)
+P_{\nu_{\bp} \rightarrow \nu_{\bp}}(\bp,\tau) (1-\prod_{\bq \ne \bp}P_{\theta_{1} \rightarrow \theta_{1}}(\bq,\tau)).
\label{BPD}
\eea
$P_S$ differs from our survival probability in Eq.(\ref{P11}) 
which corresponds to the sum of the probabilities for all the final states specified
as ${\ket {\nu_{\bp} [j_\bq]}}$,
\bea
\sum_{[j_\bq]} P_{\nu_{\bp} [j_\bq=1]\rightarrow \nu_{\bp} [j_\bq]}(\bp,\tau) &=&P_{\nu_{\bp} \rightarrow \nu_{\bp}}(\bp,\tau) \prod_{\bq \ne \bp} 
\sum_{j_{\bq}=1}^4 P_{\theta_{1} \rightarrow \theta_{j_{\bq}}}(\bq,\tau),\\
&=&P_{\nu_{\bp} \rightarrow \nu_{\bp}}(\bp,\tau) ,
\eea
where we use the unitarity of the vacuum evolution in Eq.(\ref{0toall}),
\bea
\sum_{j_{\bq}=1}^4 P_{\theta_{1} \rightarrow \theta_{j_{\bq}}}(\bq,\tau)=1.
\eea
The difference of two survival probabilities is given as,
\bea
P_{\nu_{\bp} \rightarrow \nu_{\bp}}(\bp,\tau)-P_S=P_{\nu_{\bp} \rightarrow \nu_{\bp}}(\bp,\tau) (1-\prod_{\bq \ne \bp}P_{\theta_{1} \rightarrow \theta_{1}}(\bq,\tau)). \label{Ps-P11}
\eea
The difference is exactly the same as the difference of their decay probability and our chiral oscillation probability. From Eq.(\ref{BPD}) , one can read,
\bea
P_D-P_{\nu_{\bp} \rightarrow \nu_{\bp} \bar{\nu}_{\bp} \bar{\nu}_{-\bp}}(\bp,\tau)
=P_{\nu_{\bp} \rightarrow \nu_{\bp}}(\bp,\tau) (1-\prod_{\bq \ne \bp}P_{\theta_{1} \rightarrow \theta_{1}}(\bq,\tau)).\label{Pd-P1M1}
\eea
Thus the unitarity relation holds for the two sums,
\bea
P_S+P_D=P_{\nu_{\bp} \rightarrow \nu_{\bp}}(\bp,\tau)+P_{\nu_{\bp} \rightarrow \nu_{\bp} \bar{\nu}_{\bp} \bar{\nu}_{-\bp}}(\bp,\tau)=1.
\eea

Next, we follow their method of the perturbative expansion and apply it to our oscillation probabilities.
Blasone et al. \cite{Blasone:2024zsn} employ the approximation, i.e., the perturbation with respect to $m_{e \mu}$, the off-diagonal element of the Dirac mass matrix and compute the probabilities $P_D$ and $P_S-1$ up to the second order of $m_{ e\mu}$. 
The correponding approximation to the present problem is the perturbation with respect to the Majorana mass $\frac{m}{E_\bp}$. 
Below we obtain the perturbative expression for $P_D$ and $P_S-1$ for present case up to the second order of $\frac{m}{E_\bp}$.
First the exact formulae for the relevant probabilities are shown below. From Eq.(\ref{Sqcomponent}),(\ref{Pq}), (\ref{P11}), and (\ref{P1M1}), they are given by,
\bea
&& P_{\nu_{\bp} \rightarrow \nu_{\bp} \bar{\nu}_{\bp} \bar{\nu}_{-\bp}}(\bp,\tau)=g^2(\bp, \tau), \\
&& P_{\nu_{\bp} \rightarrow \nu_{\bp}}(\bp,\tau)=1-g^2(\bp ,\tau), \\
&& P_{\theta_{1} \rightarrow \theta_{1}}(\bq,\tau)=(1-g^2(\bq,\tau))^2, \\
&& P_{\theta_{1} \rightarrow \theta_{2}}(\bq,\tau)=P_{\theta_{1} \rightarrow \theta_{3}}(\bq,\tau)=g^2(\bp,\tau)(1-g^2(\bq,\tau)), \\
&& P_{\theta_{1} \rightarrow \theta_{4}}(\bq,\tau)=g^4(\bq, \tau), 
\eea
where  $g(\bq, \tau)$ is given in Eq.(\ref{eq:g}).
We show the exact expression for $P_S$ in Eq.(\ref{BPS}) and $P_D$ in 
 Eq.(\ref{BPD}). Then we expand them up to the order of $m^2$ equivalently $g^2$.
From Eq.(\ref{BPD}), 
\bea
P_D&=&g^2(\bp,\tau) +(1-g^2(\bp,\tau))\{1-\prod_{\bq \ne \bp} (1-g^2(\bq,\tau))^2\} \label{EPD} ,\\
&\simeq& g^2(\bp,\tau)+2 \sum_{\bq \ne \bp} g^2(\bq,\tau), 
\label{APD}
\eea
The first term denotes the probability for 
the process $\nu_\bp \to \nu_\bp +\bar{\nu}_\bp + \bar{\nu}_{-\bp}$.
The second term represents the probabilites for the processes 
$\nu_\bp \to \nu_{\bp} + \bar{\nu}_\bq + \bar{\nu}_{-\bq}$ and $\nu_\bp  \to \nu_{\bp} +
{\nu}_\bq + {\nu}_{-\bq}$ where the integration over the momenta $\bq$  of final states is included explicitly. 
The survival probabilty is also given as,
\bea
P_S&\simeq&1-g^2(\bp,\tau)-2\sum_{\bq \ne \bp} g^2(\bq,\tau).
\label{APS}
\eea
As we can see from the approximate formula for the decay 
probability, Eq.(\ref{APD}) and the survival probabilty, Eq.(\ref{APS}),
both probabilities include the momentum integral,
\bea
2 \sum_{\bq \ne \bp} g^2(\bq, \tau) = 2 m^2 \sum_{\bq\ne \bp} \frac{\sin^2{E(\bq)\tau}}{E(\bq)^2}.
\eea
In their final expression of the decay probability in Eq.(51) of \cite{Blasone:2024zsn}, these terms with momentum integration were subtracted.
This contribution with momentum integration corresponds to exactly the difference between the decay probability and the chiral oscillation probability in Eq.(\ref{Pd-P1M1})
and also the difference between their survival probability and ours in Eq.(\ref{Ps-P11}),
\bea
P_{\nu_{\bp} \rightarrow \nu_{\bp}}(\bp,\tau) (1-\prod_{\bq \ne \bp}P_{\theta_{1} \rightarrow \theta_{1}}(\bq,\tau))\simeq 
2 \sum_{\bq \ne \bp} g^2(\bq, \tau).
\eea
In contrast to $P_D$ and $P_S$, the contribution with momentum integration is absent in the probilities in Eqs.(\ref{P11}, \ref{P1M1}).  
Due to the unitarity 
of the vacuum evolution, our suvirval probability and chiral oscillation probability are finite and the subtraction of the divergence is not needed.
For $P_D$ and $P_S$, the contribution with momentum integration appears and must be subtracted to obtain the finite results.  After the subtraction, 
$P_D$ and $P_S$ are in agreement with our chiral oscillation probability in Eq.(\ref{P1M1}) and survival probability in Eq.(\ref{P11}) respectively.
Therefore after subtraction of the terms with momentum integration, the third term in Eq.(\ref{APS}) and the second term in Eq.(\ref{APD}), the unitarity relation becomes
identical in both approaches.
\section{Proof of $P_{\nu_\bp \to \bar{\nu}_\bp}(\bp,\tau) =0$ in the mass basis} \label{sec:appendix proof}
In this appendix, we prove that probability for neutrino and anti-neutrino oscillation vanishes ($P_{\nu_\bp \to \bar{\nu}_\bp}(\bp,\tau) =0$) in the mass basis. In terms of the operators with definite lepton numbers, one can easily show that  $P_{\nu_\bp \to \bar{\nu}_\bp}(\bp,\tau) =0$. However, it is necessary to show that the same result can be obtained in the mass basis. $P_{\nu_\bp \to \bar{\nu}_\bp}(\bp,\tau)$ is given by,
\begin{align}
    P_{\nu_\bp \to \bar{\nu}_\bp}(\bp,\tau) = \left| \, _\bp \langle0, t_f| \beta(\bp,t_f)\alpha^\dagger(\bp,t_i) |0, t_i\rangle_{\bp} \right|^2.
\end{align}
Without loss of generality, one can set the initial time as $t_i=0$ and then $\tau=t_f-t_i=t$.  In this case, the transition amplitude $A(\nu_\bp \to \bar{\nu}_\bp)$ is,
\begin{align}
    A(\nu_\bp \to \bar{\nu}_\bp) =  \, _\bp \langle0, t| \beta(\bp,t)\alpha^\dagger(\bp,0) |0, 0\rangle_{\bp} .
\end{align}
To find this transition amplitude, the vacuum and operator are rewritten in the mass basis. First, from \cite{SalimAdam:2021suq} and \cite{Morozumi:2022mqh}, the relation between the operators with definite the lepton numbers ($\alpha(\pm \bp, t), \beta( \pm \bp, t)$) and the operators in the mass basis ($\alpha_{M}( \pm \bp,\lambda) , \lambda = \pm$) is expressed as, 
\begin{align}
 \alpha(\pm \bp, t)&= \sqrt{\frac{E_\bp+|\bp|}{2 E_\bp}}\left(\alpha_{M}( \pm \bp,-) e^{-i E_\bp t} \pm \frac{i m}{E_\bp+|\bp|} \alpha_{M}^{\dagger}(\mp \bp,-) e^{i E_\bp t}\right), \nn \\
&=\cos \phi_{p} a_{M}(\pm\bp,-)e^{-i E_\bp t} \pm i \sin \phi_{p} a_{M}^{\dagger}(\mp\bp,-)e^{i E_\bp t}, \label{ap}\\
 \beta( \pm \bp, t)&= \sqrt{\frac{E_\bp+|\bp|}{2 E_\bp}}\left(\alpha_{M}( \pm \bp,+) e^{-i E_\bp t} \pm \frac{i m}{E_\bp+|\bp|} \alpha_{M }^{\dagger}(\mp \bp,+) e^{i E_\bp t}\right) ,\nn \\ 
&=\cos \phi_{p} \alpha_{M}(\pm\bp,+)e^{-i E_\bp t} \pm i \sin \phi_{p} \alpha_{M}^{\dagger}(\mp\bp,+)e^{i E_\bp t}, \label{bp}
\end{align}
where $\phi_p$ is the momentum dependent angle and the velocity $v$ is $v = \cos 2 \phi_{p} = \frac{|\bp|}{E_\bp}$. The operator in the mass basis $a_{M}( \pm \bp,\lambda)$ satisfy the anti-commutation relations,
\begin{align}
\{\alpha_M(\bp,\lambda),\alpha_M^\dagger(\bq,\lambda')\} = \delta_{\bp\bq}\delta_{\lambda \lambda'}.
\end{align}
Next, the relation between the vacuum with definite the lepton numbers $\ket{0,t}_\bp$ and the vacuum in mass basis ${\ket{0_M}}_\bp$ is obtained from \cite{Morozumi:2022mqh} by,
\begin{align}
 |0, t\rangle_{\bp} &= \left[\cos ^{2} \phi_{p}-\sin ^{2} \phi_{p}e^{-4 i E_{p} t} B_{M+}^{\dagger}(\mathbf{p}) B_{M-}^{\dagger}(\mathbf{p})+i \sin \phi_{p} \cos \phi_{p}e^{-2 i E_{p} t} \sum_{\lambda= \pm} B_{M \lambda}^{\dagger}(\mathbf{p})\right]\ket{0_{M}}_\bp,\label{0m}
\end{align}
where $B_{M \lambda}^{\dagger}(\bp) = \alpha_{M}^{\dagger}(-\bp,\lambda)\alpha_{M}^{\dagger}(\bp,\lambda)$. The vaccum for the mass basis satisfies,
\bea
\alpha_M(\pm \bp, \pm){\ket{0_M}}_\bp=0.
\eea
From Eq.(\ref{ap}), (\ref{bp}) and (\ref{0m}), we are ready to show an initial state with a definite lepton number +1 as follows,
\begin{align}
    \alpha^\dagger(\bp,0) |0, 0\rangle_{\bp} 
    &=\left(\cos \phi_{p} \alpha_{M}^{\dagger}(\bp,-)-i \sin \phi_{p} \alpha_{M}(-\bp,-)\right)|0, 0\rangle_{\bp}, \nn\\
&= \left(\cos \phi_{p} \alpha_{M}^{\dagger}(\bp,-)-i \sin \phi_{p} \alpha_{M}(-\bp,-)\right), \nn \\
    &\times \left[\cos ^{2} \phi_{p}-\sin ^{2} \phi_{p} B_{M+}^{\dagger}(\mathbf{p}) B_{M-}^{\dagger}(\mathbf{p})+i \sin \phi_{p} \cos \phi_{p} \sum_{\lambda= \pm} B_{M \lambda}^{\dagger}(\mathbf{p})\right] {\ket{0_M}}_\bp, \nn\\
&=\left\{\cos \phi_{p}+i \sin \phi_{p} B_{M+}^{\dagger}(\bp)\right\} \alpha_{M}^{\dagger}(\bp,-){\ket{0_M}}_\bp.
\end{align}
Similarly, the final state with a definite lepton number $-1$ is shown as,
\begin{align}
    \, _\bp \langle0, t| \beta(\bp,t) &= \, _\bp \bra{0_M}\alpha_{M}(\bp,+)  e^{i E_{p} t}\left(\cos \phi_{p}-i \sin \phi_{p} e^{2 i E_{p} t} B_{M-}(\bp)\right) .
\end{align}
Thus, the transition amplitude $A(\nu(\bp)\to \bar{\nu}(\bp))$ is given by,
\begin{align}
    A(\nu_\bp \to \bar{\nu}_\bp) &= \, _\bp  \bra{0_M}\alpha_{M}(\bp,+) e^{i E_{p} t}\left(\cos \phi_{p}-i \sin \phi_{p} e^{2 i E_{p} t} B_{M-}(\bp)\right) \nn\\
    &\quad\times \left\{\cos \phi_{p} +i \sin \phi_{p} B_{M+}^{\dagger}(\bp) \right\}\alpha_{M}^{\dagger}(\bp,-){\ket{0_M}}_\bp, \\
        &=e^{i E_{p} t}\cos^2 \phi_{p } \{ \, _\bp \bra{0_M}\alpha_{M}(\bp,+) \alpha_{M}^{\dagger}(\bp,-){\ket{0_M}}_\bp \} \nn\\
    &\quad-e^{i E_{p} t}i\sin \phi_{p}\cos \phi_{p} \{\, _\bp  \bra{0_M}\alpha_{M}^\dagger(-\bp,+)\alpha_{M}^{\dagger}(\bp,-){\ket{0_M}}_\bp \} ,\nn\\
    &\quad-e^{3i E_{p} t}i\sin \phi_{p}\cos \phi_{p} \{ \, _\bp \bra{0_M}\alpha_{M}(\bp,+)\alpha_{M}(-\bp,-){\ket{0_M}}_\bp \} \nn\\
    &\quad+e^{3i E_{p} t}\sin^2 \phi_{p} \{\, _\bp \bra{0_M}\alpha_{M}^\dagger(-\bp,+)\alpha_{M}(-\bp,-){\ket{0_M}}_\bp \} \nn \\
    &=0.
\end{align}
From the above, we were able to show that neutrino and anti-neutrino oscillation also vanishes in the mass basis as $P_{\nu_\bp \to \bar{\nu}_\bp}(\bp,\tau) =0$. 
\end{appendices}

%
\end{document}